\documentclass[preprint]{ptephy}

\preprintnumber{KYUSHU-HET-185}

\usepackage{amssymb}
\usepackage{amsthm}
\usepackage{amsmath}
\usepackage{booktabs}
\usepackage{bbm}
\usepackage{mathtools}
\usepackage{subcaption}
\usepackage{braket}
\DeclareMathOperator{\tr}{tr}

\newcommand{\Slash}[1]{{\ooalign{\hfil/\hfil\crcr$#1$}}}
\numberwithin{equation}{section}

\newcommand{\be}{\begin{equation}}
\newcommand{\ee}{\end{equation}}
\newcommand{\fn}{\footnote}
\newcommand{\lt}{\left}
\newcommand{\rt}{\right}
\newcommand{\non}{\nonumber \\}
\usepackage{url}

\begin{document}

\title{Renormalon-free definition of the gluon condensate within the large-$\beta_0$ approximation}

\author{%
\name{\fname{Hiroshi} \surname{Suzuki}}{1,\ast}
and
\name{\fname{Hiromasa} \surname{Takaura}}{1}
}

\address{%
\affil{1}{Department of Physics, Kyushu University
744 Motooka, Nishi-ku, Fukuoka, 819-0395, Japan}
\email{hsuzuki@phys.kyushu-u.ac.jp}
}

\date{\today}

\begin{abstract}
We propose a clear definition of the gluon condensate within the large-$\beta_0$ approximation as an attempt
toward a systematic argument on the gluon condensate.
We define the gluon condensate such that it is free from a renormalon uncertainty,
consistent with the renormalization scale independence of each term of the operator product expansion (OPE),
and an identical object irrespective of observables.
The renormalon uncertainty of $\mathcal{O}(\Lambda^4)$, 
which renders the gluon condensate ambiguous, is separated from a perturbative calculation 
by using a recently suggested analytic formulation.
The renormalon uncertainty is absorbed into the gluon condensate in the OPE, which makes
the gluon condensate free from the renormalon uncertainty.
As a result, we can define the OPE in a renormalon-free way.
Based on this renormalon-free OPE formula, we discuss numerical extraction of the gluon condensate 
using the lattice data of the energy density operator defined by the Yang--Mills gradient flow.
\end{abstract}
\subjectindex{B01, B06, B32, B65}

\maketitle


\section{Introduction}
\label{sec:1}
In perturbative expansion of observables in quantum chromodynamics (QCD),
the coefficients of the series typically grow factorially as a function of the order 
and thus the perturbation series is an asymptotic series at
best~\cite{LeGuillou:1990nq}. One of the origins of this growth is the factorial
increase in the number of Feynman diagrams with respect to the order. In the
renormalized perturbation theory, there is another origin: there exists a class
of Feynman diagrams whose \emph{amplitude\/} grows
factorially~\cite{tHooft:1977xjm,LeGuillou:1990nq, Beneke:1998ui}. This kind of factorial
behavior produces the so-called renormalon ambiguity in perturbation theory of
order~$e^{-4\pi u/(\beta_0\alpha)}\sim(\Lambda^2/Q^2)^u$, where $\beta_0$ is the
one-loop coefficient of the beta function, $\alpha$ the renormalized coupling,
constant and $u$~parametrizes the ``strength'' of the renormalon; $\Lambda$ is
the renormalization group invariant mass scale and $Q$ is a typical energy scale
in the problem under consideration. ($\beta_0=\frac{11}{3} C_A-\frac{4}{3} T_F n_f>0$ in our convention.) 
Particularly for the (dimensionless) observables which are Lorentz invariant and dependent on a single energy scale,
perturbative calculations suffer from the so-called $u=2$ renormalon, 
and have the inevitable uncertainty of $\mathcal{O}(\Lambda^4/Q^4)$. 
Examples of such observables are the Adler function, the plaquette, the energy density operator defined by the Yang--Mills gradient flow, etc.\fn{
The static QCD potential at very short distances also suffers from the $u=2$ renormalon, although it is not Lorentz invariant.}

The operator product expansion (OPE), which is an extended framework of perturbation theory,
is considered to be helpful in overcoming the error due to the renormalon.
The OPE of a general observable $X(Q^2)$ with the above properties is of the form
\be
X(Q^2)=c_{\mathbbm{1}, X}(Q^2) \langle \mathbbm{1} \rangle +c_{FF, X}(Q^2) \frac{ \langle \frac{\alpha}{\pi} \{F_{\mu\nu}^aF_{\mu\nu}^a\}_R(x) \rangle}{Q^4}
   +\mathcal{O}(Q^{-6}) \, , \label{OPE}
\ee
in quenched QCD.
Here, the coefficients $c_{\mathbbm{1},X}$ and $c_{FF, X}$ denote the Wilson coefficients,
and the symbol $R$ stands for renormalization.  
(We can adopt, for instance, the $\overline{\rm MS}$ scheme to define renormalized composite operators.)
The Wilson coefficients are calculated in perturbation theory,
whereas the vacuum expectation values (VEVs) of composite operators are generally nonperturbative objects.
In particular, the VEV of $\frac{\alpha}{\pi} \{F_{\mu\nu}^aF_{\mu\nu}^a\}_R(x)$ is known as the gluon condensate.
(These condensates are zero in perturbative calculations in dimensional regularization.)
Hence, the Wilson coefficient $c_{\mathbbm{1}, X}$ is given by perturbative calculation of~$X(Q^2)$
and possesses the renormalon uncertainty of $\mathcal{O}(\Lambda^4/Q^4)$.
This error is the same order of magnitude as the second term of the OPE, 
the first nonperturbative effect specified by the gluon condensate.
Hence, the gluon condensate has been considered as a key element to overcome 
the error due to the renormalon. 
In particular, since the gluon condensate appears universally in the OPE and conceptually has a unique value irrespective of observables,
determining this value (in some way) would be quite helpful; 
it allows us to predict $\mathcal{O}(\Lambda/Q)^4$ terms of many observables.

However, in order to determine the gluon condensate numerically in the context of the OPE, 
one cannot avoid the issue of how to deal with the renormalon uncertainty in $c_{\mathbbm{1},X}(Q^2)$.
In fact, the gluon condensate cannot be determined in the following naive treatment.
From the OPE \eqref{OPE}, the gluon condensate is read off from the coefficient of the $1/Q^4$ term in 
$[X(Q^2)-c_{\mathbbm{1}, X}(Q^2)]/c_{FF, X}(Q^2)$ while measuring an observable $X(Q^2)$ nonperturbatively  (for instance using lattice).
However, since $c_{\mathbbm{1}, X}(Q^2)$ has an error of $\mathcal{O}(\Lambda^4/Q^4)$,
the determined gluon condensate has an error of $\mathcal{O}(\Lambda^4)$, which is the same size as the gluon condensate itself.
We note that the renormalon uncertainty is the minimum error of perturbation theory.
Thus, this argument indicates that the gluon condensate has a significant error 
{\it {even when}} one has sufficiently large-order results. 



There have been some proposals concerning treatment of $c_{\mathbbm{1}, X}(Q^2)$ to extract the gluon condensate   
\cite{Lee:2010hd, Bali:2014sja, Lee:2015bci, Bali:2015cxa, DelDebbio:2018ftu} (see also Ref.~\cite{Horsley:2012ra}). 
An often adopted prescription is to use $c_{\mathbbm{1},X}(Q^2)$
that is obtained by truncating the perturbative series at the $n_*$th order 
where the $n_*$th order term is minimal among the terms in the perturbative series.
However, the following properties are not assured in this prescription:
(i) each term in the OPE is independent of the renormalization scale, and 
(ii) the gluon condensate is a universal and identical object irrespective of observables.
Regarding the first issue, the truncation order $n_*$ varies depending on the renormalization scale 
since it is  given by $n_* \sim \frac{8 \pi}{\beta_0 \alpha(\mu)}$.
It is explicitly shown in Ref.~\cite{Mishima:2016vna} that, in the so-called large-$\beta_0$ approximation, 
a different choice of renormalization scale indeed changes the truncated result of  $c_{\mathbbm{1}, X}(Q^2)$.
This indicates that $c_{\mathbbm{1}}$ is {\it dependent} on the renormalization scale  and so is the second term,
which contradicts the property usually used that each term of the OPE is {\it independent} of the renormalization scale.\footnote{
An appropriate redefinition of the renormalized operator $\frac{\alpha}{\pi} \{ F_{\mu\nu}^aF_{\mu\nu}^a\}_R(x)$ 
and $c_{FF}(Q)$ can make each of them renormalization scale independent at all order since the operator is proportional 
to the trace part of the energy--momentum tensor, which is renormalization scale independent.
This issue is not relevant to the present argument because the problem is whether the combination of 
$c_{FF}(Q) \langle \frac{\alpha}{\pi} \{ F_{\mu\nu}^aF_{\mu\nu}^a\}_R(x) \rangle$, which is independent of the redefinition,
is renormalization scale dependent or not. We also note that 
the renormalized operator $\frac{\alpha}{\pi} \{ F_{\mu\nu}^aF_{\mu\nu}^a\}_R(x)$ is renormalization scale independent
at the one-loop level.}
In addition to this, the gluon condensate defined in this way has not been shown to be identical
to the ones defined from other observables. 
If property (ii) is not assured, an extracted value of the gluon condensate from an observable 
has a very limited meaning:
it cannot be used as an input in the OPE \eqref{OPE} of other observables.

In this paper, using the large-$\beta_0$ approximation \cite{Beneke:1994qe,Broadhurst:1993ru,Ball:1995ni}, we propose a definition of the gluon condensate
which explicitly satisfies (i) and (ii). That is, our definition of the gluon condensate is compatible with
the renormalization scale independence of each term of the OPE
and is unique irrespective of observables.
Thus, it qualifies as an input to the $\mathcal{O}(\Lambda^4/Q^4)$ term of the OPE of broad observables.
Also, it  does not suffer from the renormalon uncertainty of $c_{\mathbbm{1}, X}(Q^2)$.

We achieve this as follows. We regularize the all-order perturbative series of $c_{\mathbbm{1}, X}(Q^2)$
by introducing an infrared (IR) cutoff scale $\mu_f$.
Following Refs.~\cite{Mishima:2016vna, Mishima:2016xuj}, we separate this regularized Wilson coefficient $c_{\mathbbm{1},X}(Q^2; \mu_f)$
 into its cutoff-dependent and -independent parts,
which correspond to the renormalon uncertainty and renormalon-independent (renormalon-free) parts, respectively.
The renormalon-free part becomes the first term in our OPE.
On the other hand, the renormalon uncertainty of $\mathcal{O}(\mu_f^4/Q^4)$ \fn{This renormalon uncertainty corresponds
to the $\mathcal{O}(\Lambda^4/Q^4)$ renormalon uncertainty, which one encounters in a regularization without using the IR cutoff.}
is absorbed into the second term of the OPE.
It will be shown for some explicit observables that the renormalon uncertainty of the gluon condensate
(which is exhibited as the ultraviolet (UV) cutoff dependence) is exactly canceled by this procedure.
In other words, each term of our OPE (up to the second term) can be defined as a renormalon-free object.
In particular, the second term of our OPE is specified by the renormalon-free gluon condensate
whose definition is explicitly given in this paper.
In this construction, each term of the OPE is also independent of the renormalization scale (that is different from the cutoff scale).
This is realized because the first term of our OPE (the renormalon-free part of $c_{\mathbbm{1},X}$)
is obtained based on the all-order perturbative series, which is renormalization scale independent.
Moreover, the gluon condensate defined in this paper is observable independent,
which is related to the universality of the renormalon cancelation.

We note that in deriving these features, the large-$\beta_0$ approximation is always assumed.
At this stage, it is not obvious how the relations and formulas presented in this paper
are modified beyond this approximation.
Also, since the large-$\beta_0$ approximation is accurate only at the leading logarithmic level,
it is difficult to obtain some physical consequences, such as the value of the defined gluon condensate,
from comparison of our theoretical result with the lattice result.
(To determine the gluon condensate precisely, 
we at least have to know the large-order perturbative behavior as already mentioned.)

Nevertheless, we believe that the present work makes an improvement in our conceptual understanding
of the gluon condensate because we can explicitly show how the gluon condensate is made well defined.
It is also notable that the large-$\beta_0$ approximation can simulate well the divergent behavior of perturbative series
caused by renormalons. 
We thus expect that the present work provides a foundation
to define a gluon condensate with good nature [(i) and (ii)] in a more systematic approach
beyond this approximation.

The paper is organized as follows.
In Sect.~\ref{sec:2}, we give a definition of the renormalon-free gluon condensate, which is based on the $u=2$ renormalon cancelation in the OPE.
In this section, we treat general observables with the $u=2$ renormalon as the first IR renormalon.
In~Sect.~\ref{sec:3}, we study some examples and confirm the renormalon cancelation explicitly.
A main example is the energy density operator defined by the Yang--Mills gradient flow.
The conclusions and discussion are given in Sect.~\ref{sec:4}.
In Appendix~\ref{sec:A}, we collect our notational conventions.
In Appendix~\ref{sec:B}, we explain construction of the large-$\beta_0$ approximation in the context of the gradient flow.
In Appendix~\ref{sec:C}, we compare the perturbative series of the energy density operator defined
by the Yang--Mills gradient flow obtained in the exact calculations and in the large-$\beta_0$ approximation.
In Appendix~\ref{sec:D}, we report an attempt at a numerical determination of the gluon condensate, applying the formula presented in this paper.

\section{Renormalon-free definition of the gluon condensate} 
\label{sec:2}
To define the gluon condensate unambiguously in the OPE, 
it is necessary to separate the associated renormalon uncertainty from the Wilson coefficient $c_{\mathbbm{1}, X}(Q^2)$ in Eq.~\eqref{OPE}.
For this, we use the formulation proposed in Ref.~\cite{Mishima:2016vna},
a review of which is given in Sect.~\ref{subsec:2.1}.
In Sect.~\ref{subsec:2.2}, we present a definition of the renormalon-free gluon condensate in light of the renormalon cancelation,
and in Sect.~\ref{subsec:2.3}, we consider the scheme dependence of a renormalon-free gluon condensate.

\subsection{Formula to separate the renormalon in $c_{\mathbbm{1}, X}$}
\label{subsec:2.1}
We consider a Euclidean dimensionless observable $X(Q^2)$ which depends on a single scale $Q$
and has the first IR renormalon at $u=2$.
Let us assume that the leading-order (LO) term of $X(Q^2)$ in perturbation theory is $\mathcal{O}(\alpha)$
and is given by a one-gluon exchanging diagram. 
This is the case, for instance, for 
the Adler function\footnote{We study the reduced Adler function, where $\mathcal{O}(\alpha^0)$ term is subtracted.} and the energy density operator defined by the Yang--Mills gradient flow. 
For such observables, we can construct all-order perturbative series in the so-called large-$\beta_0$ approximation
\cite{Beneke:1994qe,Broadhurst:1993ru,Ball:1995ni}.

\begin{figure}[t]
\centering
\includegraphics[width=7cm]{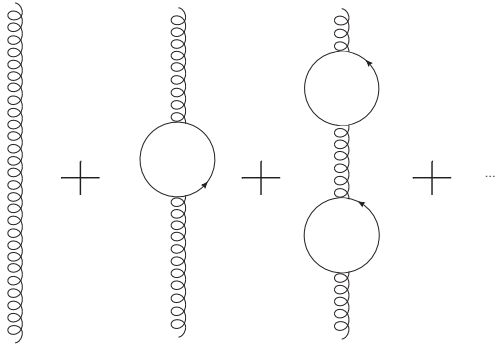}
\caption{The infinite sum of fermion loop chains.}
\label{fig:1}
\end{figure}

The construction is as follows (see also Appendix~\ref{sec:B}).
We consider insertion of a chain of fermion bubbles into the gluon propagator of the LO diagram; see Fig.~\ref{fig:1}.
Each bubble produces a factor proportional to $\frac{\alpha(\mu)}{4\pi} \frac{4}{3} T_F n_f \log(e^{5/3} \mu^2/p^2)$,
where $n_f$ is the number of flavors, $\mu$ a renormalization scale, and $p$ the gluon momentum. 
In Appendix~\ref{sec:A}, we present our convention for the normalization factors.
In the large-$\beta_0$ approximation, we replace $-\frac{4}{3} T_F n_f \to  \beta_0$, where\footnote{We
define the beta function as~$\mu^2d\alpha/(d\mu^2)\equiv\beta(\alpha)
=-(\beta_0/4\pi)\alpha^2+\mathcal{O}(\alpha^3)$, where
$\alpha\equiv g^2/(4\pi)$.}
\be
\beta_0\equiv\frac{11}{3}C_A-\frac{4}{3} n_f T_F \, ,
\ee
and then obtain the series as
\be
X(Q^2)_{\rm pert}=c_{\mathbbm{1}, X}(Q^2)
=\sum_{n=0}^{\infty} \int \frac{d^4 p}{(2 \pi)^4} F_X(p,Q) \alpha(\mu) \lt[\frac{\beta_0 \alpha(\mu)}{4 \pi} \log(e^{5/3} \mu^2/p^2) \rt]^n \, .
\ee
The function $F_X(p,Q)$ is the integrand determined from the LO diagram.
In the first equality, we use the fact that the perturbative series of $X(Q^2)$
coincides with that of $c_{\mathbbm{1}, X}(Q^2)$ in the context of the OPE. 
This is because the condensates in Eq.~\eqref{OPE} are nonperturbative objects and 
zero in perturbative evaluation (with dimensional regularization).
We note that before the replacement $-\frac{4}{3} T_F n_f \to  \beta_0$ the series gives the leading contribution in the large-$n_f$ limit.
However, the large-$\beta_0$ approximation obtained after this replacement is not justified in any limit of the QCD parameters.
Nevertheless, this series gives the exact leading-logarithmic (LL) contribution 
of perturbative series. In addition, it is empirically known that this series
gives a good approximation of the first few to several terms which have been calculated explicitly.

The series in the large-$\beta_0$ approximation can be resummed and expressed as 
\be
c_{\mathbbm{1},X}(Q^2)=\int_0^{\infty} \frac{d \tau}{2 \pi \tau} w_X(\tau/Q^2) \alpha_{\beta_0}(\tau) \, , \label{c1X}
\ee
where $\tau$ is the modulus of the gluon momentum $\tau=p^2$;
$w_X(\tau/Q^2)$ is a dimensionless function originating from $F_X(p,Q)$ \fn{
An explicit relation between $F_X(p,Q)$ and $w_X(\tau/Q^2)$ is given by
\be
w_X(\tau/Q^2)=\frac{\tau^2}{2} \int  \frac{d \Omega_3^p}{(2 \pi)^3} F_X(p,Q) \bigg|_{|p^2|=\tau} \nonumber
\ee
where $d^4p=d |p| d \Omega_3^p |p|^3$.
} and depends on a single variable $\tau/Q^2$;
and $\alpha_{\beta_0}(\tau)$ is the running coupling specific in the large-$\beta_0$ approximation, 
\be
\alpha_{\beta_0}(\tau)=\sum_{n=0}^{\infty} \alpha(\mu) \lt[\frac{\beta_0 \alpha(\mu)}{4 \pi} \log(e^{5/3} \mu^2/\tau^2) \rt]^n
=\frac{4 \pi}{\beta_0} \frac{1}{\log \lt(\frac{\tau}{e^{5/3} \Lambda^2} \rt)} \, .
\ee
Here, we used the expression of the one-loop running coupling $\alpha(\mu)=\frac{4  \pi}{\beta_0} \frac{1}{\log(\mu^2/\Lambda^2)}$,
where $\Lambda$ is a renormalization group independent scale: $\Lambda^2=\mu^2 e^{-4 \pi /(\beta_0 \alpha(\mu))}$.
We note that Eq.~\eqref{c1X} is independent of the renormalization scale.

Equation~\eqref{c1X} is just formal because the integrand has a single pole on the integration path at $\tau=e^{5/3} \Lambda^2$. 
We regularize this quantity with an IR cutoff scale $\mu_f$:
\be
c_{\mathbbm{1},X}(Q^2;\mu_f)=\int_{\mu_f^2}^{\infty} \frac{d \tau}{2 \pi \tau} w_X(\tau/Q^2) \alpha_{\beta_0}(\tau) \, , \label{c1Xcut}
\ee
where $\Lambda \ll \mu_f \ll Q$.
This resummed quantity explicitly depends on the regularization parameter (the cutoff scale).
This feature that the resummation depends on how to be regularized is common in the presence of IR renormalons.
In this formulation, the IR renormalons are related to the function $w_X$: 
the IR renormalons determine the expansion of $w_X(x)$ in $x$ \cite{Neubert:1994vb}. 
In particular, the $u=2$ renormalon as the first IR renormalon leads to
\be
w_X(x)=b_{2, X} x^2+({\text{higher-order terms in $x$}}) \, , \label{wexp}
\ee
where $b_{2, X}$ is an ($x$-independent) constant.
Thus, the cutoff dependence (dependence on $\mu_f/Q$) of Eq.~\eqref{c1Xcut} is determined by the IR renormalons. 
In this sense, the cutoff dependence corresponds to the renormalon uncertainty.
On the other hand, a cutoff-independent part, which potentially exists, corresponds to the renormalon-free part.

\begin{figure}[htbp]
\centering
\begin{subfigure}{0.45\columnwidth}
\centering
\includegraphics[width=0.7\columnwidth]{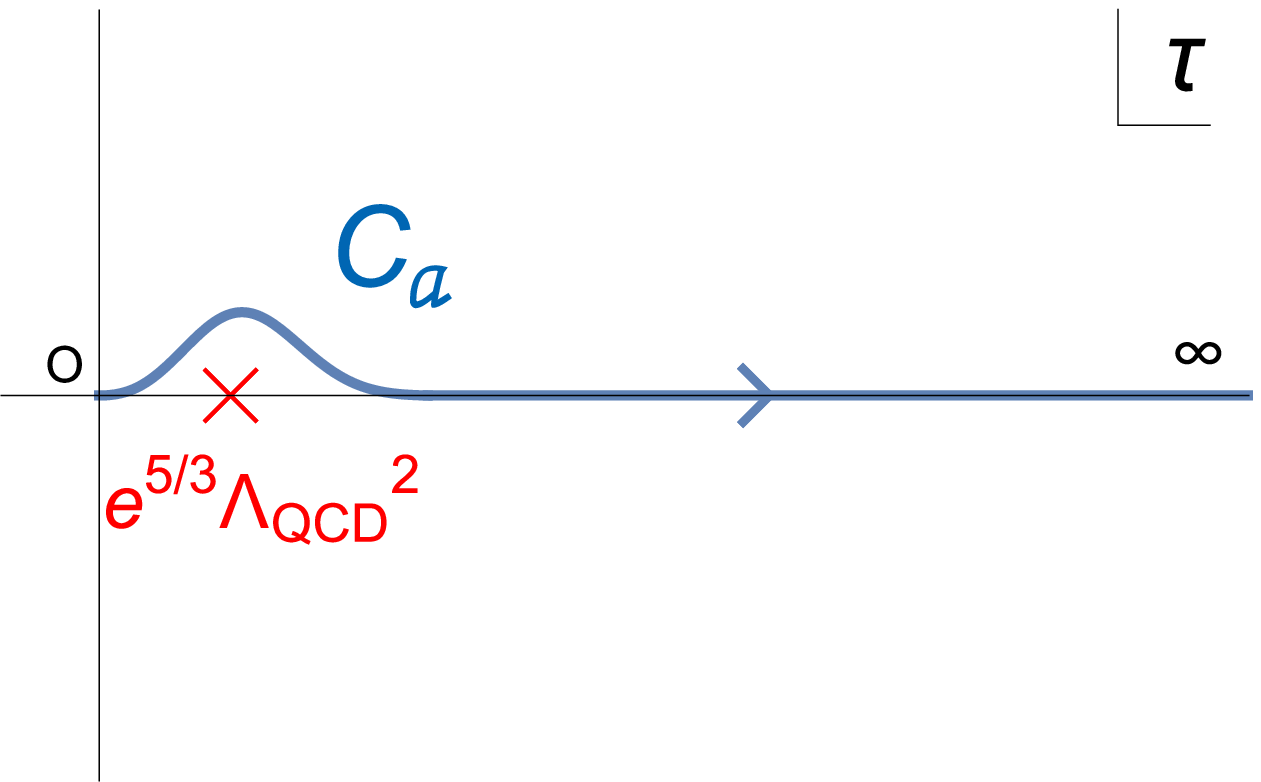}
\end{subfigure}
\hspace*{2em}
\begin{subfigure}{0.45\columnwidth}
\centering
\includegraphics[width=0.7\columnwidth]{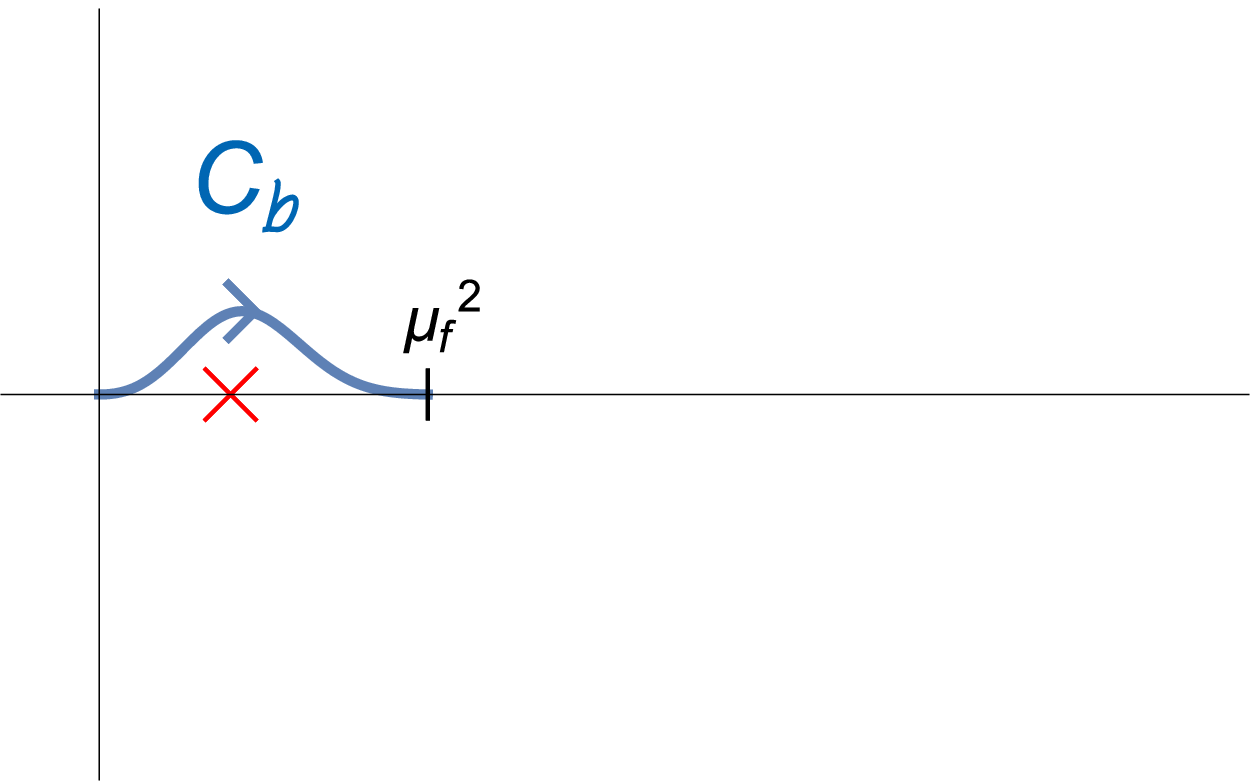}
\end{subfigure}
\caption{The integration contours $C_a$ and~$C_b$ used
in~Eq.~\eqref{CaCb}.}
\label{fig:2}
\end{figure}

This motivates us to extract the cutoff-independent part from Eq.~\eqref{c1Xcut},
which is precisely calculated in a renormalon-free way.
This is carried out by
(I) rewriting the integrand by a new analytic function $W_X(z)$ 
defined in the complex $z$-plane and satisfying $2 \, {\rm Im} W_X(z)=w_X(z)$ for $z \in \mathbb{R}_{\geq 0}$, 
and then (II) deforming the integration contour in the complex $\tau$ plane.
The function $W_X(z)$ can be constructed by
\be
W_X(z)=\int_0^{\infty} \frac{d x}{2 \pi} \frac{w_X(x)}{x-z-i0} \, . \label{WX}
\ee
With this function we can rewrite Eq.~\eqref{c1Xcut} as
\be
c_{\mathbbm{1},X}(Q^2; \mu_f)={\rm Im} \lt(\int_{C_a}-\int_{C_b} \rt) \frac{d \tau}{\pi \tau} \, W_X(\tau/Q^2) \alpha_{\beta_0}(\tau) \, , \label{CaCb}
\ee
where the integration contours $C_a$ and $C_b$ are displayed in Fig.~\ref{fig:2}.

The integral along $C_a$ is obviously independent of $\mu_f$.
Actually, we also obtain $\mu_f$-independent parts from the integral along $C_b$. 
In this integral, we first expand $W_X(z)$ in $z$:
\be
W_X(z)=a_{0,X}+a_{1,X} z+\lt(a_{2,X}(z)+ i \frac{b_{2,X}}{2} \rt) z^2 +\cdots \, .
\ee
As a generic feature with the first IR renormalon at $u=2$,
the coefficients of the $z^0$ and $z^1$~terms are real 
whereas the coefficient of the $z^2$ term is complex.
($a_{2,X}(z)$ is a polynomial of $\log {z}$.)
This follows from Eq.~\eqref{wexp} and $2 \, {\rm Im} W_X(z)=w_X(z)$ for $z \in \mathbb{R}_{\geq 0}$.
For a real part (i.e., a term with coefficient $a_{n,X}$), the integral is evaluated as
\begin{align}
{\rm Im} \int_{C_b} \frac{d \tau}{\pi \tau} \, a_{n,X}(\tau/Q^2) \lt(\frac{\tau}{Q^2} \rt)^n \alpha_{\beta_0}(\tau)
&=\frac{1}{2 i} \lt({\int}_{C_b}-\int_{C_b^*} \rt) \frac{d \tau}{\pi \tau} \, a_{n,X}(\tau/Q^2) \lt(\frac{\tau}{Q^2} \rt)^n \alpha_{\beta_0}(\tau) \non
&=-\frac{4 \pi a_{n,X}(e^{5/3} \Lambda^2/Q^2)}{\beta_0} \lt( \frac{e^{5/3} \Lambda^2}{Q^2}\rt)^n \, .
\end{align}
In the first equality, we used a property of the integrand $\{f(z)\}^*=f(z^*)$.
Thus, the integration path can be deformed to a circle surrounding the pole,
which yields the $\mu_f$-independent result.
On the other hand, the integral of the $b_{2,X}$-term, 
\be
{\rm Im} \int_{C_b} \frac{d \tau}{2 \pi \tau} \, i b_{2,X} \lt(\frac{\tau}{Q^2} \rt)^2 \alpha_{\beta_0}(\tau)
\ee
remains $\mu_f$ dependent.
In this way, we have separated the $\mu_f$-independent part $c_{\mathbbm{1}, X}^{\rm RF}(Q)$, which is the renormalon-free part, from the $\mu_f$-dependent part:
\be
c_{\mathbbm{1}, X}(Q^2;\mu_f)=c_{\mathbbm{1}, X}^{\rm RF}(Q^2)-{\rm Im} \int_{C_b} \frac{d \tau}{2 \pi \tau} i b_{2,X} \lt(\frac{\tau}{Q^2} \rt)^2 \alpha_{\beta_0}(\tau)+\mathcal{O}(1/Q^6) \, ,\label{sepa}
\ee
where $c_{\mathbbm{1}, X}^{\rm RF}(Q^2)$ consists of all the $\mu_f$-independent contributions [up to $\mathcal{O}(\Lambda^4/Q^4)$]: 
\begin{align}
c_{\mathbbm{1}, X}^{\rm RF}(Q^2)&=
\lt[{\rm Im} \int_{C_a} \frac{d \tau}{\pi \tau} W_X(\tau/Q^2) \alpha_{\beta_0}(\tau)+\frac{4 \pi a_{0,X}}{\beta_0} \rt] \non
&\qquad{}
+\frac{4 \pi a_{1,X}}{\beta_0} \frac{e^{5/3} \Lambda^2}{Q^2}+\frac{4 \pi a_{2,X}(e^{5/3} \Lambda^2/Q^2)}{\beta_0} \lt( \frac{e^{5/3} \Lambda^2}{Q^2}\rt)^2 \, . \label{c1RF}
\end{align}
In Eq.~\eqref{sepa}, the first term, $c_{\mathbbm{1}, X}^{\rm RF}(Q^2)$, is $\mu_f$ independent and its asymptotic form is $\sim \alpha(Q)$~\cite{Mishima:2016vna}.
The second term is $\mu_f$ dependent and represents the leading $\mu_f$ dependence of $c_{\mathbbm{1},X}(Q^2;\mu_f)$ as $\mathcal{O}(\mu_f^4/Q^4)$.\fn{
This inevitable uncertainty of order $\mu_f^4/Q^4$ corresponds to the $\mathcal{O}(\Lambda^4/Q^4)$ uncertainty 
which one encounters in the resummation using the Borel integral where the integration contour is deformed as $\int_{0}^{\infty} \to \int_{0 \pm i \epsilon}^{\infty\pm i \epsilon}$.} 
Thus, the first term gives a dominant contribution at large $Q$ due to $1/Q^2 \sim e^{-4 \pi/(\beta_0 \alpha(Q))}$.
Hence, Eq.~\eqref{sepa} can be regarded as an expansion in $1/Q$.
The last term of $\mathcal{O}(1/Q^6)$ generally has both cutoff-independent and -dependent terms, 
but dose not play any role in the following discussion.

Applying this formulation, one can calculate $c_{\mathbbm{1},X}^{\rm RF}(Q^2)$ for explicit observables. 
In particular, once the function $w_X$ for the observable under consideration is obtained, 
one can follow the above calculations.
As an example, we will study the energy density operator defined by the Yang--Mills gradient flow in Sect.~\ref{sec:3}.

\subsection{Renormalon-free definition of the gluon condensate in light of renormalon cancelation}
\label{subsec:2.2}
Let us consider the relation between the Wilson coefficient we have calculated [$c_{\mathbbm{1},X}(Q^2;\mu_f)$] and the OPE.
We have regularized the Wilson coefficient $c_{\mathbbm{1},X}(Q^2)$ with the IR cutoff scale $\mu_f$.
This implies that UV contributions are calculated as the perturbative contribution.
Accordingly, it is natural that the remaining mode below the cutoff scale is represented by the nonperturbative contributions.\footnote{
This is analogous to the integration-by-regions argument \cite{Beneke:1997zp,Smirnov:1999bza}, 
where hard contributions in loop integrals are identified with Wilson coefficients
whereas soft contributions correspond to condensates.}
Hence, we  introduce $\mu_f$ as the {\it UV} cutoff scale to the condensates in the OPE.
Thus, we perform the OPE as
\be
X(Q^2)=c_{\mathbbm{1},X}(Q^2;\mu_f)+c_{FF, X} \frac{ \langle \frac{\alpha}{\pi} \{F_{\mu\nu}^aF_{\mu\nu}^a\}_R(x; \mu_f) \rangle}{Q^4}+\mathcal{O}(1/Q^6) \, , \label{mufOPE}
\ee
where the gluon condensate possesses the UV cutoff scale $\mu_f$.
For $c_{FF, X}$, we eliminate the argument $Q$ since it is a $Q$-independent constant under the approximation we consider.\fn{
As seen from the explicit calculation in Eq.~\eqref{mufglu} below, $\langle \frac{\alpha}{\pi} \{F_{\mu\nu}^aF_{\mu\nu}^a\}_R(x; \mu_f) \rangle$ 
has the same order of magnitude as $c_{\mathbbm{1}, X}(Q^2;\mu_f)$ in the large-$\beta_0$ approximation. (They are $\mathcal{O}(\alpha)$.)
Since in the OPE~Eq.~\eqref{mufOPE}, each term has the same order of magnitude in the large-$\beta_0$ approximation, 
the coefficient $c_{FF,X}$, which is calculated in perturbation theory, 
is thus $\mathcal{O}(\alpha^0)$ and does not have $Q$-dependence. 
(We note that $c_{FF, X}$ is renormalization scale independent at the one-loop level.)}

In Eq.~\eqref{mufOPE}, the cutoff dependence should be canceled in the sum of the first and second terms
because the observable is independent of the cutoff.
Remember that the cutoff dependence of the first term represents the renormalon uncertainty.
Hence, such a cancelation corresponds to the renormalon cancelation in the OPE.
If this is true, the $\mu_f$ dependence of the second term of Eq.~\eqref{mufOPE} should be given by
\be
c_{FF,X} \frac{ \langle \frac{\alpha}{\pi} \{F_{\mu\nu}^aF_{\mu\nu}^a\}_R(x; \mu_f) \rangle}{Q^4} 
\overset{\mu_f \, {\rm dep.}}{\sim} 
\int_{0}^{\mu_f^2} \frac{d \tau}{2 \pi \tau} b_{2,X} \lt(\frac{\tau}{Q^2} \rt)^2 \alpha_{\beta_0}(\tau)  \label{correspondence}
\ee
since the cutoff dependence is canceled in the quantity
\be
\int_0^{\infty} \frac{d \tau}{2 \pi \tau} w_X(\tau/Q^2) \alpha_{\beta_0}(\tau)
=\lt(\int_{\mu_f^2}^{\infty}+ \int_{0}^{\mu_f^2} \rt)\frac{d \tau}{2 \pi \tau} w_X(\tau/Q^2) \alpha_{\beta_0}(\tau) \, , 
\ee
and $c_{\mathbbm{1}, X}(Q;\mu_f)$ is defined as the first integral [cf. Eq.~\eqref{c1Xcut}].
In Eq.~\eqref{correspondence}, we used the expansion of $w_X(\tau/Q^2)$ given in Eq.~\eqref{wexp}.
In fact, Eq.~\eqref{correspondence} can be reduced to the relation between the coefficients $c_{FF,X}$ and $b_{2, X}$.
To see this, we calculate the gluon condensate in the large-$\beta_0$ approximation with the UV cutoff.
In calculating this local product, we use a naive point-splitting regularization and then contract the gauge fields.
It reads
\be
\langle \frac{\alpha}{\pi} \{F_{\mu\nu}^a F_{\mu\nu}^a\}_R(x; \mu_f) \rangle
\overset{\mu_f \, {\rm dep.}}{\sim}   A \int_0^{\mu_f^2} \frac{d \tau}{\pi \tau} \, \tau^2 \alpha_{\beta_0}(\tau) \, , \label{mufglu}
\ee
with
\be
A=\frac{3 \, {\rm dim}(G)}{8 \pi^2} \, . \label{A}
\ee
This cutoff dependence is regarded as the renormalon uncertainty of the gluon condensate.\fn{
We believe that the {\it cutoff dependence} of the gluon condensate can be calculated in perturbation theory due to $\mu_f \gg \Lambda$.
On the other hand, its exact behavior (determined by the low energy dynamics) cannot be obtained in perturbation theory 
(as the expression based on perturbation theory [right-hand side of Eq.~\eqref{mufglu}] is not well-defined).}
One sees that, using Eq.~\eqref{mufglu}, both sides of Eq.~\eqref{correspondence} have the same $\tau$ integral.
Hence, the renormalon cancelation \eqref{correspondence} requires  
\be
b_{2,X}=2 A c_{FF, X} \, . \label{candb}
\ee

We confirm the relation \eqref{candb} (equivalent to the renormalon cancelation) for explicit examples below (Sect.~\ref{sec:3}).
Thus, we use this relation in the following general argument.

We now define the renormalon-free gluon condensate. 
Using the separation formula obtained in Eq.~\eqref{sepa}, we express the OPE \eqref{mufOPE} as
\begin{align}
X(Q^2)&=
c_{\mathbbm{1}, X}^{\rm RF}(Q^2)-{\rm Im} \int_{C_b} \frac{d \tau}{2 \pi \tau} i b_{2,X} \lt(\frac{\tau}{Q^2} \rt)^2 \alpha_{\beta_0}(\tau) \non
&\qquad{}
+c_{FF, X} \frac{ \langle \frac{\alpha}{\pi} \{F_{\mu\nu}^aF_{\mu\nu}^a\}_R(x; \mu_f) \rangle}{Q^4}+\mathcal{O}(1/Q^6) \, . \label{OPEsepa}
\end{align}
Then using the relation \eqref{candb}, we obtain 
\begin{align}
X(Q^2)&=
c_{\mathbbm{1}, X}^{\rm RF}(Q^2)-c_{FF, X} {\rm Im} \int_{C_b} \frac{d \tau}{\pi \tau} i A  \lt(\frac{\tau}{Q^2} \rt)^2 \alpha_{\beta_0}(\tau) \non
&\qquad{}
+c_{FF, X} \frac{ \langle \frac{\alpha}{\pi} \{F_{\mu\nu}^aF_{\mu\nu}^a\}_R(x; \mu_f) \rangle}{Q^4}+\mathcal{O}(1/Q^6) \non
&\equiv c_{\mathbbm{1},X}^{\rm RF}(Q^2)+c_{FF, X} \frac{ \langle \frac{\alpha}{\pi} \{F_{\mu\nu}^aF_{\mu\nu}^a\}_R(x) \rangle^{\rm RF}}{Q^4}+\mathcal{O}(1/Q^6) \, . \label{RFOPE}
\end{align}
Here we make the renormalon uncertainty of $c_{\mathbbm{1}, X}(Q;\mu_f)$ absorbed into the second term
and define the renormalon-free gluon condensate as
\be
\langle \frac{\alpha}{\pi} \{F_{\mu\nu}^aF_{\mu\nu}^a\}_R(x) \rangle^{\rm RF}
\equiv \langle \frac{\alpha}{\pi} \{F_{\mu\nu}^aF_{\mu\nu}^a\}_R(x; \mu_f) \rangle -
A \, {\rm Im}  \int_{C_b} \frac{d \tau}{ \pi \tau} i    \tau^2 \alpha_{\beta_0}(\tau) \, . \label{RFglu}
\ee
We now can perform the OPE where each term is free from the renormalon uncertainty [as shown in the last expression of Eq.~\eqref{RFOPE}].

The features of our definition of the gluon condensate \eqref{RFglu} can be stated as follows.
First, it is certainly free from the renormalon (or independent of the cutoff scale) 
since the second term in Eq.~\eqref{RFglu} exhibits the opposite $\mu_f$ dependence to the first term calculated in~Eq.~\eqref{mufglu}.
Secondly, the definition does not have observable dependence.
In other words, the symbol $X$ does not appear in~Eq.~\eqref{RFglu} 
but is encoded only in its Wilson coefficient $c_{FF, X}$.
(We note that the operator $\frac{\alpha}{\pi} \{F_{\mu\nu}^aF_{\mu\nu}^a\}_R(x;\mu_f)$ is obviously observable independent
because it is a basis of the OPE taken universally for general observables.)
Thus, the gluon condensate \eqref{RFglu} is a universal quantity.
This is compatible with its original (and naive) concept.
Note that realization of this feature is not trivial a priori in the presence of the renormalon uncertainty. 
Thirdly, each term of the OPE [in the last expression of Eq.~\eqref{RFOPE}] is renormalization scale independent.
This stems from the fact that the first term $c_{\mathbbm{1}, X}^{\rm RF}(Q^2)$ is renormalization scale independent
(since it is based on the all-order perturbative series) and the observable $X(Q^2)$ is, of course, renormalization scale independent.
This is again consistent with the original OPE structure.
In this way, we realize the definition of a gluon condensate with the desired properties.

The renormalon-free gluon condensate is considered to have a nonperturbative contribution.
Thus, it is difficult to calculate this quantity theoretically. 
Instead, treating it as a fitting parameter, we extract its value from comparison of the renormalon-free OPE \eqref{RFOPE} with a measurement of $X(Q^2)$ (for instance using lattice simulations)---see Appendix~\ref{sec:D}.\fn{
The first term $c_{\mathbbm{1}, X}^{\rm RF}(Q)$ can always be calculated theoretically according to the above method.}
This quantity depends only on the dynamical scale $\Lambda$ and is independent of the regularization parameter $\mu_f$.
We again emphasize that the value of the gluon condensate is common regardless of the chosen observables.
Hence, once its value is determined from {\it an} observable, its value can be used in the renormalon-free OPE \eqref{RFOPE} 
of other observables as an input to predict the $\Lambda^4/Q^4$ terms.
Such a prediction is beyond perturbation theory because it overcomes the error of the renormalon uncertainty of $\mathcal{O}(\Lambda^4/Q^4)$.

\subsection{Conversion to other schemes}
\label{subsec:2.3}
There are potentially many schemes to define the renormalon-free gluon condensate.
In this sense, we adopt one of possible schemes.
Scheme conversion can be done by changing the identification of {\it a cutoff-independent part}. 
One can change a cutoff-independent part to
\be
c_{\mathbbm{1},X}^{\rm{RF}'}(Q^2) \equiv c_{\mathbbm{1}, X}^{{\rm RF}}(Q^2)+s_X \frac{\Lambda^4}{Q^4}  \, ,
\ee
where $s_X$ is a ($Q$-independent) constant.
Then, the OPE \eqref{OPEsepa} is rearranged in this different scheme as
\begin{align}
X(Q^2)&=
c_{\mathbbm{1}, X}^{\rm RF'}(Q^2)-s_X \frac{\Lambda^4}{Q^4} -{\rm Im} \int_{C_b} \frac{d \tau}{2 \pi \tau} i b_{2,X} \lt(\frac{\tau}{Q^2} \rt)^2 \alpha_{\beta_0}(\tau) \non
&\qquad{}
+c_{FF, X} \frac{ \langle \frac{\alpha}{\pi} \{F_{\mu\nu}^aF_{\mu\nu}^a\}_R(x; \mu_f) \rangle}{Q^4}+\mathcal{O}(1/Q^6) \non
&=c_{\mathbbm{1}, X}^{{\rm RF}'}(Q^2)+c_{FF,X} \frac{ \langle \frac{\alpha}{\pi} \{F_{\mu\nu}^aF_{\mu\nu}^a\}_R(x) \rangle^{{\rm RF}'}}{Q^4}+\mathcal{O}(1/Q^6)
\end{align}
where
\be
 \langle \frac{\alpha}{\pi} \{F_{\mu\nu}^aF_{\mu\nu}^a\}_R(x) \rangle^{{\rm RF}'}
 =\langle \frac{\alpha}{\pi} \{F_{\mu\nu}^aF_{\mu\nu}^a\}_R(x) \rangle^{{\rm RF}}-\frac{s_X}{c_{FF,X}} \Lambda^4 \, . \label{glucon2}
\ee
It is notable that in order to keep the observable-independent nature of the renormalon-free gluon condensate,
the parameter $s_X$ should be taken as $s_X=c_{FF,X} s'$, where $s'$ is an arbitrary constant and is independent of observables.
Namely, $s_X$ is not completely arbitrary but should be proportional to~$c_{FF,X}$.  
Once $s_X$ is taken in this way, the renormalon-free gluon condensate defined in this new scheme satisfies the three features stated in Sect.~\ref{subsec:2.2}.
However, as seen from this discussion, it seems quite natural to choose the scheme with~$s_X=0$, where the gluon condensate is 
obviously independent of observables.
In addition, the scheme where $s_X=0$ corresponds to a minimal subtraction of the cutoff dependence of the gluon condensate.

We note that the existence of other schemes as above is not problematic.
This is because the gluon condensate is not directly related to a physical observable 
but is a partial contribution to it.
Thus the gluon condensate can be scheme dependent in the above sense.
We also note, however, that we have now clarified the relation between different schemes as given in~Eq.~\eqref{glucon2}.
This allows us to compare the gluon condensates in different schemes systematically.

\section{Explicit examples}
\label{sec:3}
In this section, we study explicit examples: the Adler function and the energy density operator defined by the Yang--Mills gradient flow. 
We also mention previous work concerning the static QCD potential.
We explicitly confirm the renormalon cancelation \eqref{candb} for these quantities.
\\

\subsection{Adler function}
As the first example, we consider the reduced Adler function $D(Q^2)$ \cite{Adler:1974gd}.\fn{
The reduced Adler function is defined such that its perturbative expansion starts at $\mathcal{O}(\alpha)$.}
It is defined as
\be
D(Q^2)=4 \pi^2 Q^2 \frac{d \Pi(Q^2)}{d Q^2} -\frac{N}{3}
\ee
for $n_f=1$ and $G=SU(N)$, where $\Pi(Q^2)$ is a correlator of the quark current $J_{\mu}(x)=\bar{q}(x) \gamma_{\mu} q(x)$,\fn{
Although we basically consider quenched QCD in this paper, the quark field is necessary to consider the Adler function.
Here, we briefly discuss modification of our analysis for QCD with massless quarks.
In this case, the condensate of the dimension-4 operator $\langle \bar{\psi} {\Slash{D}} \psi \rangle$ can appear in the OPE.
The renormalon uncertainty of this condensate, which is exhibited by the cutoff dependence, 
can appear at $\mathcal{O}(\alpha)$
since the cutoff is introduced to the gluon momentum in our calculations.
However, the contribution at this order is zero, as shown by an explicit perturbative calculation,
where the two diagrams are canceled.
Hence, it does not show cutoff dependence at this order.
As a consequence, this condensate does not have the renormalon uncertainty in the large-$\beta_0$ approximation.
This is indeed consistent with the observation below that the renormalon uncertainty of $c_{\mathbbm{1}}$
is canceled against that of the gluon condensate alone.
We note that, however, this does not necessarily mean $\langle \bar{\psi} {\Slash{D}} \psi \rangle =0$
since we might have nonperturbative contributions.  In this sense, it would be appropriate that we add
a $\langle \bar{\psi} {\Slash{D}} \psi \rangle/Q^4$ term to the OPE of Eq.~\eqref{RFOPEAdler}. }
\be
(Q_{\mu} Q_{\nu}-\delta_{\mu \nu} Q^2) \Pi(Q^2)=\int d^4 x \, e^{iQx}  \langle T J_{\mu}(x) J_{\nu}(0) \rangle \, .
\ee
The renormalon separation for the reduced Adler function has been calculated in Refs.~\cite{Mishima:2016xuj,Mishima:2016vna}.
In particular, $c_{\mathbbm{1},D}^{\rm RF} (Q^2)$ has been explicitly obtained.
The result has the same form as~Eq.~\eqref{sepa}:
\be
c_{\mathbbm{1},D}(Q^2;\mu_f)=c_{\mathbbm{1},D}^{\rm RF}(Q^2)
-{\rm Im} \int_{C_b} \frac{d \tau}{2 \pi \tau} i b_{2,D} \lt(\frac{\tau}{Q^2} \rt)^2 \alpha_{\beta_0}(\tau)+\mathcal{O}(1/Q^6) 
\ee
where 
\be
b_{2,D}=N C_F={\rm dim} (G) T_F  \label{b2D} \, .
\ee
In the OPE [Eq.~\eqref{OPE}] for the (reduced) Adler function, the Wilson coefficient $c_{FF}^D$ is given by \cite{Lee:2011te}
\be
c_{FF, D}=\frac{4 \pi^2}{3} T_F \, . \label{cFFD}
\ee
The results in Eqs.~\eqref{b2D} and \eqref{cFFD} indeed indicate the renormalon cancelation \eqref{candb}.
Hence, we can perform the OPE as
\be
D(Q^2)=c_{\mathbbm{1},D}^{\rm RF}(Q^2)+c_{FF,D} \frac{\langle \frac{\alpha}{\pi} \{ F^a_{\mu \nu} F^a_{\mu \nu} \}_R \rangle^{\rm RF}}{Q^4}+\mathcal{O}(1/Q^6) \, , \label{RFOPEAdler}
\ee
where the renormalon-free gluon condensate \eqref{RFglu} properly emerges.
\\

\subsection{Energy density operator in the Yang--Mills gradient flow} 
As the second example, 
we investigate the energy density operator defined by the Yang--Mills gradient flow 
(which is denoted by $\hat{E}(t^{-1})$ below),
where the typical scale is $Q^2=t^{-1}$. ($t$ is the flow time as explained shortly.)
We first extract the renormalon-free part $c_{\mathbbm{1} ,\hat{E}}(t^{-1})$ using the method in Sect.~\ref{subsec:2.1}
and then examine the renormalon cancelation.

The Yang--Mills gradient flow \cite{Narayanan:2006rf, Luscher:2010iy} is a one-parameter evolution of the gauge
field~$A_\mu(x)$ defined by the flow equation,\footnote{Our notational
convention is summarized in~Appendix~\ref{sec:A}. The term that is proportional
to the ``gauge-fixing parameter''~$\alpha_0$ in~Eq.~\eqref{eq:(1.1)} is
introduced to simplify the perturbative argument on the gauge degrees of
freedom. Although this term breaks the gauge covariance, it can be shown that
any gauge-invariant quantity is independent of~$\alpha_0$. This gauge-breaking
term is thus physically irrelevant.}
\begin{equation}
   \partial_tB_\mu(t,x)=D_\nu G_{\nu\mu}(t,x)
   +\alpha_0D_\mu\partial_\nu B_\nu(t,x),\qquad
   B_\mu(t=0,x)=A_\mu(x) \, .
\label{eq:(1.1)}
\end{equation}
$t\geq0$~is called the flow time, where dim$[t]=-2$; 
$B_{\mu}(t, x)$ is the flowed gauge field and coincides with $A_{\mu}(x)$ at $t=0$;
$G_{\mu\nu}(t,x)$ is the field
strength of the flowed gauge field~$B_\mu(t,x)$,
\begin{equation}
   G_{\mu\nu}(t,x)
   =\partial_\mu B_\nu(t,x)-\partial_\nu B_\mu(t,x)
   +g_0 [B_\mu(t,x),B_\nu(t,x)] \, ,
\label{eq:(1.2)}
\end{equation}
and the covariant derivative is also defined with respect to~$B_\mu(t,x)$,
\begin{equation}
   D_\mu=\partial_\mu+g_0 [B_\mu,\cdot] \, .
\label{eq:(1.3)}
\end{equation}

We define the energy density operator as
\begin{equation}
   E(t,x)\equiv\frac{g_0^2}{4}G_{\mu\nu}^a(t,x)G_{\mu\nu}^a(t,x) \, .
\label{eq:(1.4)}
\end{equation}
As the renormalizability theorem~\cite{Luscher:2011bx} implies, 
its VEV is a renormalized finite quantity although it is a certain combination of the
\emph{bare\/} gauge fields through the flow equation.\footnote{Reference~\cite{Hieda:2016xpq} is an exposition on the
renormalizability theorem.} 
Thus, this quantity can be regarded as a physical observable 
and is quite useful for the scale setting and the non-perturbative definition of the gauge coupling in
the context of lattice gauge theory; see the review~\cite{Ramos:2015dla}, and the
recent paper~\cite{Brida:2016flw} and the references cited therein.
In the following, we study the dimensionless energy density operator given by
\begin{equation}
  \hat{E}(t^{-1})
   \equiv t^2 \left\langle  E(t,x)\right\rangle \, .
\label{eq:(1.5)}
\end{equation}

We calculate $\hat{E}(t^{-1})$, in particular its Wilson coefficient of the identity operator in the small flow time expansion (analog of the OPE)
in the large-$\beta_0$ approximation.
In~Appendix~\ref{sec:B}, we explain how to apply the large-$\beta_0$ approximation in the gradient flow formalism.
From Eq.~\eqref{Elargebeta}, the Wilson coefficient of the identity operator for $\hat{E}$ is obtained as
\be
c_{\mathbbm{1},\hat{E}}(t^{-1}; \mu_f)=\int_{\mu_f^2}^{\infty} \frac{d \tau}{2 \pi \tau} w_{\hat{E}}(\tau t) \alpha_{\beta_0}(\tau)
\ee
with
\be
w_{\hat{E}}(x)=2 A \pi^2  x^2 e^{-2 x} \, , \label{wE}
\ee
where the constant $A$ is given by Eq.~\eqref{A}. To extract the renormalon-free part, we construct~$W_{\hat{E}}$ [cf. Eq.~\eqref{WX}].
For convenience, we present $W_{\hat{E}+}(z) \equiv W_{\hat{E}}(-z)$,
which has no singularities for a real positive~$z$:
\be
W_{\hat{E}+}(z)=A \pi \lt(\frac{1}{4}-\frac{z}{2}+e^{2 z} z^2 \Gamma(0,2z) \rt) \, ,
\ee
where $\mathit{\Gamma}(a,z)\equiv\int_z^\infty dt\,t^{a-1}e^{-t}$ is the
incomplete Gamma function.
According to the method in Sect.~\ref{subsec:2.1}, we can construct the renormalon-free part through the function $W_{\hat{E}}(z)$ and its expansion in $z$,
\be
W_{\hat{E}}(z)=A \pi \lt[\frac{1}{4}+\frac{1}{2} z-(\gamma_{E}+\log{2}+\log{z}-i \pi) z^2+\cdots \rt] \, . \label{WEexp}
\ee
We then obtain
\be
c_{\mathbbm{1},\hat{E}}(t^{-1}; \mu_f)=c_{\mathbbm{1},\hat{E}}^{\rm RF}(t^{-1})-{\rm Im} \int_{C_b} \frac{d \tau}{2 \pi \tau} i b_{2, \hat{E}} (\tau t)^2 \alpha_{\beta_0}(\tau)
+\mathcal{O}(t^3) \label{Esepa}
\ee
with
\begin{align}
c_{\mathbbm{1}, \hat{E}}^{\rm RF}(t^{-1})
&=\lt[\int_0^{\infty} \frac{d \tau}{\pi \tau} W_{\hat{E}+}(t \tau) {\rm Im} \, \alpha_{\beta_0}(-\tau+i0)+\frac{A \pi^2}{\beta_0}\rt] \non
 &\qquad{}
+\frac{2 A \pi^2}{\beta_0} e^{5/3} t \Lambda^2-\frac{4 \pi^2 A (\gamma_{E}+\log{2}+\log({e^{5/3} t \Lambda^2}) )}{\beta_0} (e^{5/3} t \Lambda^2)^2 \, .  \label{ERF}
\end{align}
$c_{\mathbbm{1},\hat{E}}^{\rm RF}(t^{-1})$ is obtained from the general result \eqref{c1RF},
but for the integral along $C_a$ we deform the integral path $C_a$ into $\tau=0 \to - \infty$ 
using a good convergence property of $W_{\hat{E}}(z)$ at~$|z| \to \infty$.
In~Fig.~3, we plot the renormalon-free part
$c_{\mathbbm{1}, {\hat{E}}}^{{\rm RF}}$ for $G=SU(3)$ (and~$n_f=0$).
\begin{figure}[tbp]
\centering
\includegraphics[width=0.55\columnwidth]{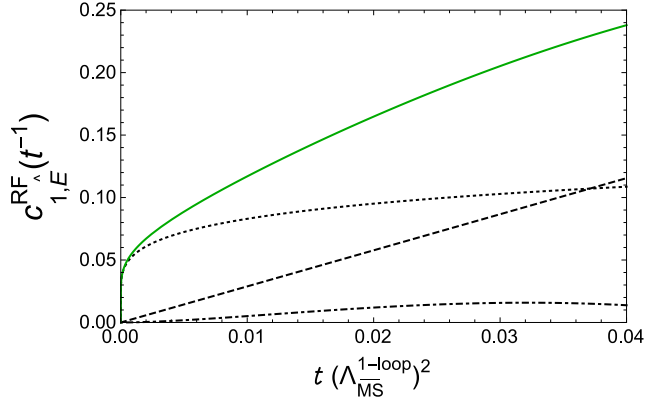}
\caption{The renormalon-free part $c_{\mathbbm{1},\hat{E}}^{\rm RF}(t^{-1})$~[Eq.~\eqref{ERF}] for $G=SU(3)$ and~$n_f=0$ (green solid line). 
We also plot each term of~Eq.~\eqref{ERF}:
the first term (inside square brackets) is shown by the black dotted line, 
the second term of $\mathcal{O} (t\Lambda^2)$ by the black dashed
line, and the last $\mathcal{O}((t\Lambda^2)^2)$ term by the black dot-dashed line.}
\label{fig:3}
\end{figure}

Let us confirm the renormalon cancelation. In Eq.~\eqref{Esepa}, $b_{2, \hat{E}}$ is given by
\be
b_{2,\hat{E}}=2  A \pi^2 \, , \label{b2E}
\ee
which is read off from the expansion of Eq.~\eqref{wE} or \eqref{WEexp}.
In the OPE [for $\hat{E}(t)$ one should regard $Q^2=t^{-1}$ in Eq.~\eqref{OPE}],
the Wilson coefficient $c_{FF,\hat{E}}$ is given by
\be
c_{FF,\hat{E}}=\pi^2 \label{cFFE}
\ee
due to $G^a_{\mu \nu} G^a_{\mu \nu} \sim F^a_{\mu \nu} F^a_{\mu \nu}$ at the tree level (after subtracting $c_{\mathbbm{1},\hat{E}}$).\fn{
The Wilson coefficient of this operator has been calculated at NLO in Refs.~\cite{Suzuki:2013gza,Makino:2014taa}.}
These results [Eqs.~\eqref{b2E} and \eqref{cFFE}] indicate the renormalon cancelation \eqref{candb}.
Hence, we can perform the OPE in a renormalon-free way:
\be
\hat{E}(t^{-1})=c_{\mathbbm{1},\hat{E}}^{\rm RF}(t^{-1})+c_{FF,\hat{E}} \langle \frac{\alpha}{\pi} \{F^a_{\mu \nu} F^a_{\mu \nu} \}_R\rangle^{\rm RF} t^2+\mathcal{O}(t^3) \, , \label{EtOPE}
\ee
where the renormalon-free gluon condensate appears.
\\

\subsection{Static QCD potential}
The static QCD potential at very short distances has the first IR renormalon at $u=2$.
The above renormalon separation has been carried out in Ref.~\cite{Takaura:2017lwd}, and as a result, 
the renormalon-free gluon condensate \eqref{RFglu} has been shown to appear in its OPE.

\section{Conclusions and discussion}
\label{sec:4}

In this paper, we have given a clear definition of the gluon condensate.
It is given in the context of the OPE of the observables whose perturbative predictions suffer from
the $\mathcal{O}(\Lambda^4)$ ($u=2$) renormalon uncertainty.
The definition of the gluon condensate is closely related to the issue of how to treat the renormalon uncertainty
of the Wilson coefficient of the identity operator, which is the first term of the OPE. (For perturbative evaluation,
we used the large-$\beta_0$ approximation.)
In our formulation, we separated the renormalon uncertainty of the Wilson coefficient from the renormalon-free part
using a recently suggested analytic formula.
The renormalon-free part is the first term of our OPE,
while the renormalon uncertainty is absorbed into the second term described by the gluon condensate.
It was explicitly shown for some examples that by this procedure the renormalon uncertainty of the gluon condensate
(which is exhibited by the UV cutoff dependence) is canceled.
We defined this renormalon-free quantity as the gluon condensate.
It has the following desired properties:
it is free from the renormalon uncertainty of $\mathcal{O}(\Lambda^4)$,
consistent with the renormalization scale invariance of each term of the OPE,
and an identical object irrespective of observables.
Thus our definition is free from various instabilities, while the above properties are not always assured 
in previously adopted definitions of the gluon condensate in the literature.

Explicit advantages of the above definition can be stated as follows.
First, the renormalon-free gluon condensate is independent of the artificially introduced parameter (namely, the cutoff scale)
and is dependent only on the dynamical scale $\Lambda$. Thus, it would be a proper
quantity to detect the low-energy dynamics of QCD.
Secondly, since it is defined as a universal quantity regardless of chosen observables,
it has a unique value. Therefore, once the value is extracted from the renormalon-free OPE formula of {\it an} observable,
it can be used as an input to predict the $\mathcal{O}(\Lambda^4)$ term of other (and many) observables.
Hence, such a formulation is quite useful to overcome the renormalon problem
that the $\mathcal{O}(\Lambda^4)$ term of observables cannot be predicted in perturbation theory.

As a main example in this paper, we studied the energy density operator defined by the Yang--Mills gradient flow.
We investigated its renormalon structure (Appendix~\ref{sec:B}) and extracted its renormalon-free part (Sect.~\ref{sec:3}).
We also discussed a numerical determination of the defined gluon condensate using the lattice data of this quantity (Appendix~\ref{sec:D}).

We remark that our results and discussion are all based on the large-$\beta_0$ approximation.
Since the large-$\beta_0$ approximation is accurate only at the leading logarithmic level, 
it is required to further develop this framework in order to realize a more realistic and preferable definition of the gluon condensate.
Indeed, the current framework is shown to be insufficient at a practical level
 as discussed in Appendix~\ref{sec:D}, where we attempt to determine the gluon condensate numerically using lattice data
of the energy density operator defined by the Yang--Mills gradient flow.
Nevertheless, the present work demonstrated how the gluon condensate can be a theoretically well-defined quantity
in the large-$\beta_0$ approximation, which can simulate the renormalon divergence of perturbative series in QCD qualitatively.
We believe that this knowledge promotes theoretical understanding on the renormalon uncertainty, the gluon condensate, and the OPE.
We also expect that this work provides a foundation for constructing a more systematic framework 
beyond the large-$\beta_0$ approximation.\footnote{For the static QCD
potential, renormalon subtraction has been carried out beyond the large-$\beta_0$ approximation~\cite{Sumino:2005cq}.
} We hope to come back to this issue in the near future.

\section*{Acknowledgments}
We are grateful to Masakiyo Kitazawa for providing us with the lattice data and
discussion.
The works of H.S. and H.T. are supported in part by JSPS Grants-in-Aid for Scientific
Research nos. JP16H03982 and JP19K14711, respectively.

\appendix

\section{Notational convention}
\label{sec:A}
We set the normalization of anti-Hermitian generators~$T^a$ of the
representation~$R$ of the gauge group~$G$ as~$\tr_R(T^aT^b)=-T_R \delta^{ab}$
and~$T^aT^a=-C_R \mathbbm{1}$. We denote $\tr_R(1)=\dim(R)$. From the
structure constants defined by~$[T^a,T^b]=f^{abc}T^c$, we set
$f^{acd}f^{bcd}=C_A \delta^{ab}$. For example, for the fundamental
$N$~representation of~$G=SU(N)$ for which $\dim(N)=N$, our normalization is
\begin{equation}
   C_A=N,\qquad T_F=\frac{1}{2},\qquad
   C_F=\frac{N^2-1}{2N}.
\label{eq:(A1)}
\end{equation}

The $D$-dimensional Euclidean action of the vectorial gauge theory is given by
\begin{equation}
   S=\int\mathrm{d}^Dx\, \frac{1}{4} F_{\mu\nu}^a(x)F_{\mu\nu}^a(x)
   +\int\mathrm{d}^Dx\,\Bar{\psi}(x) \Slash{D}\psi(x) \, .
\label{eq:(A2)}
\end{equation}
The field strength is defined by
\begin{equation}
   F_{\mu\nu}(x)
   =\partial_\mu A_\nu(x)-\partial_\nu A_\mu(x)+g_0 [A_\mu(x),A_\nu(x)],
\label{eq:(A3)}
\end{equation}
for $A_\mu(x)=A_\mu^a(x)T^a$ and~$F_{\mu\nu}(x)=F_{\mu\nu}^a(x)T^a$, 
where $g_0$ is the bare gauge coupling. The covariant derivative on the fermion is
\begin{equation}
   D_\mu=\partial_\mu+g_0 A_\mu,
\label{eq:(A4)}
\end{equation}
and $\Slash{D}\equiv\gamma_\mu D_\mu$, where $\gamma_\mu$ denotes the Hermitian
Dirac matrix.

\section{Large-$\beta_0$ approximation in the Yang--Mills gradient flow}
\label{sec:B}
We explain how to calculate $\hat{E}(t^{-1})$ [given in Eq.~\eqref{eq:(1.5)}] in the large-$\beta_0$ approximation.
First, to extract a gauge-invariant subset of Feynman diagrams that gives
the renormalon, we consider the large-$n_f$ approximation ($n_f\gg1$), while $g_0^2n_f$~is
held fixed; $g_0$~is the bare gauge coupling. With our notational convention
in~Appendix~\ref{sec:A}, the bare propagator of the gauge field in the large-$n_f$ approximation is
given by\footnote{
We adopt dimensional regularization in which the spacetime
dimension is set to be~$D\equiv4-2\epsilon$. We also use the abbreviation,
\begin{equation}
   \int_p\equiv\int\frac{\mathrm{d}^Dp}{(2\pi)^D}.
\label{eq:(2.1)}
\end{equation}
}
\begin{equation}
   \left\langle g_0^2 A_\mu^a(x)A_\nu^b(y)\right\rangle
   =g_0^2 \delta^{ab}\int_pe^{ip(x-y)}\frac{1}{(p^2)^2}
   \left\{(p^2\delta_{\mu\nu}-p_\mu p_\nu)\left[1-\omega(p)\right]^{-1}
   +\frac{1}{\lambda_0}p_\mu p_\nu\right\},
\label{eq:(2.2)}
\end{equation}
where $\lambda_0$~is the bare gauge-fixing parameter.
Note that the insertion of the fermion vacuum polarization into the gluon propagator 
is not suppressed but its contribution is $\mathcal{O}(n_f^0)$.
Hence, in this expression, we have the factor~$[1-\omega(p)]^{-1}$ 
arising from the geometric sum of fermion loop
chains in~Fig.~\ref{fig:1}, where $\omega(p)$ is the vacuum polarization given
by
\begin{equation}
   \omega(p)=\frac{1}{16\pi^2}g_0^2(4\pi e^{-\gamma_E})^\epsilon
   (p^2)^{-\epsilon}\left(\frac{1}{\epsilon}+\frac{5}{3}\right)
   \left(-\frac{4}{3}T_R n_f\right),
\label{eq:(2.3)}
\end{equation}
and $\gamma_E$ is the Euler constant.
From Eq.~\eqref{eq:(2.3)}, we see that the
renormalization in the $\overline{\text{MS}}$ scheme is accomplished by
\begin{equation}
   g_0^2=g^2\mu^{2\epsilon}(4\pi e^{-\gamma_E})^{-\epsilon}\mathcal{Z}^{-1},\qquad
   \lambda_0=\lambda\mathcal{Z}^{-1},\qquad
   \mathcal{Z}
   =1+\frac{1}{\epsilon}\frac{1}{16\pi^2}g^2
   \left(-\frac{4}{3}T_R n_f\right).
\label{eq:(2.4)}
\end{equation}
Then the renormalized gauge field propagator at leading order in the large-$n_f$ approximation is given by
\begin{align}
   &\left\langle g_0^2 A_\mu^a(x)A_\nu^b(y)\right\rangle
\notag\\
   &=g^2 \delta^{ab}\int_pe^{ip(x-y)}
\notag\\
   &\qquad{}
   \times\frac{1}{(p^2)^2}
   \left\{(p^2\delta_{\mu\nu}-p_\mu p_\nu)
   \left[1-\frac{1}{16\pi^2}g^2
   \left(-\frac{4}{3}T_R n_f\right)
   \ln\left(\frac{e^{5/3}\mu^2}{p^2}\right)\right]^{-1}
   +\frac{1}{\lambda}p_\mu p_\nu\right\} \, .
\label{eq:(2.5)}
\end{align}
Note that there is no need of the wave function renormalization of the gauge
field at the order we consider. 

The large-$n_f$ approximation can also be considered for correlation functions
of the flowed gauge fields defined by~Eq.~\eqref{eq:(1.1)}. The formal solution
of~Eq.~\eqref{eq:(1.1)} is given by~\cite{Luscher:2010iy}
\begin{equation}
   B_\mu(t,x)
   =\int\mathrm{d}^Dy\left[
   K_t(x-y)_{\mu\nu}A_\nu(y)
   +\int_0^t\mathrm{d}s\,K_{t-s}(x-y)_{\mu\nu}R_\nu(s,y)
   \right],
\label{eq:(2.6)}
\end{equation}
where
\begin{equation}
   K_t(x)_{\mu\nu}=\int_p\frac{e^{ipx}}{p^2}
   \left[(\delta_{\mu\nu}p^2-p_\mu p_\nu) e^{-tp^2}
   +p_\mu p_\nu e^{-\alpha_0tp^2}\right]
\label{eq:(2.7)}
\end{equation}
is the heat kernel and
\begin{equation}
   R_\mu=g_0 (2[B_\nu,\partial_\nu B_\mu]
   -[B_\nu,\partial_\mu B_\nu]
   +(\alpha_0-1)[B_\mu,\partial_\nu B_\nu]
   +g_0 [B_\nu,[B_\nu,B_\mu]])
\label{eq:(2.8)}
\end{equation}
represents non-linear terms in the flow equation~\eqref{eq:(1.1)}. Then, by
iteratively solving~Eq.~\eqref{eq:(2.6)}, we have a perturbative expansion of
the flowed field~$B_\mu(t,x)$ in terms of the initial value~$A_\nu(y)$. A
correlation function of the flowed gauge fields~$B$ in perturbation theory is
then computed as a correlation function of~$A$. 
In particular, the leading flowed gauge field
propagator~$\langle g_0^2 B_\mu^a(t,x)B_\nu^b(s,y)\rangle$ is given, after the
substitutions $B_\mu^a(t,x)=\int d^Dz\,K_t(x-z)_{\mu\rho}A_\rho^a(z)$
and~$B_\nu^b(t,y)=\int d^Dw\,K_t(y-w)_{\nu\sigma}A_\sigma^b(w)$, by contracting
$A_\rho^a(z)$ and~$A_\sigma^b(w)$ by~Eq.~\eqref{eq:(2.2)}; the contribution of
the non-linear term~$R_\mu$~\eqref{eq:(2.8)} always lowers the power of~$n_f$. 
In this way, the flowed gauge field
propagator in the large-$n_f$ approximation is given by
\begin{align}
   &\left\langle g_0^2 B_\mu^a(t,x)B_\nu^b(s,y)\right\rangle
\notag\\
   &=g^2 \delta^{ab}\int_pe^{ip(x-y)}
\notag\\
   &\qquad{}
   \times\frac{1}{(p^2)^2}
   \Biggl\{(p^2\delta_{\mu\nu}-p_\mu p_\nu)
   \left[1-\frac{1}{16\pi^2}g^2
   \left(-\frac{4}{3}T_R n_f\right)
   \ln\left(\frac{e^{5/3}\mu^2}{p^2}\right)\right]^{-1}e^{-(t+s)p^2}
\notag\\
   &\qquad\qquad\qquad\quad{}
   +\frac{1}{\lambda}p_\mu p_\nu e^{-\alpha_0(t+s)p^2}\biggr\}.
\label{eq:(2.9)}
\end{align}
The parameter $\alpha_0$ does not receive the
renormalization~\cite{Luscher:2011bx,Hieda:2016xpq}. The large-$n_f$ expression
of~$\hat{E}(t^{-1})$~\eqref{eq:(1.5)} is then simply given by contracting
two gauge fields in~$E(t,x)$ by the propagator~\eqref{eq:(2.9)} (it is easy to
see that the other Feynman diagrams that potentially contribute
to~$\hat{E}(t)$ always lower the powers of~$n_f$). The contraction
yields
\begin{equation}
   \hat{E}(t^{-1})
   =\frac{3\dim(G)g^2}{2}
  t^2 \int_pe^{-2tp^2}\frac{1}
   {1-\frac{1}{16\pi^2}g^2\left(-\frac{4}{3}T_R n_f\right)
   \ln\left(\frac{e^{5/3}\mu^2}{p^2}\right)}.
\label{eq:(2.10)}
\end{equation}
This is the expression in the leading order of the large-$n_f$ approximation.

Now, the large-$\beta_0$ approximation is simply defined by replacing the factor~$-\frac{4}{3}T_R n_f$ in the above
expression by the one-loop coefficient of the beta function,
\begin{equation}
   -\frac{4}{3}T_R n_f\to\beta_0\equiv\frac{11}{3}C_A-\frac{4}{3}T_R n_f.
\label{eq:(2.11)}
\end{equation}
That is, in this large-$\beta_0$ approximation, $\hat{E}(t^{-1})$ is given
by
\begin{align}
  \hat{E}(t^{-1})
   &=\frac{3\dim(G)}{2}4\pi\alpha
   t^2\int_pe^{-2tp^2}\frac{1}
   {1-\frac{\alpha}{4\pi}\beta_0
   \ln\left(\frac{e^{5/3}\mu^2}{p^2}\right)}
   \notag\\
   &=\alpha
   \int_0^\infty \frac{d\tau}{2 \pi \tau}\, 2 \pi^2 A  (t \tau)^2 e^{-2t\tau}\frac{1}
   {1-\frac{\alpha}{4\pi}\beta_0
   \ln\left(\frac{e^{5/3}\mu^2}{\tau}\right)}\notag\\
   &=\int_0^\infty \frac{d\tau}{2 \pi \tau}\, 2 \pi^2 A  (t \tau)^2 e^{-2t\tau}\alpha_{\beta_0}(\tau) \, ,
\label{Elargebeta}
\end{align}
where we have set~$\tau\equiv p^2$ and 
used $A$ defined in Eq.~\eqref{A}.

By expanding~Eq.~\eqref{Elargebeta} with respect to~$\alpha$, we have the
perturbative series in the large-$\beta_0$ approximation,
\begin{equation}
  \hat{E}(t^{-1})
   \sim \frac{3 {\rm dim} (G)}{32 \pi} \alpha\sum_{n=0}^\infty\Tilde{k}_n\alpha^n,\qquad
   \Tilde{k}_n=4 \int_0^\infty dx \,x e^{-2x}
   \left(\frac{\beta_0}{4\pi}\right)^n
   \ln^n\left(\frac{e^{5/3} t \mu^2}{x}\right),
\label{eq:(2.14)}
\end{equation}
where $\Tilde{k}_0=1$ agrees with the exact LO calculation. 
We also compare the first two perturbative coefficients with the exact perturbative
coefficients obtained in~Refs.~\cite{Luscher:2010iy,Harlander:2016vzb} in Appendix~\ref{sec:C}.

The Borel transform corresponding to the perturbative
series~\eqref{eq:(2.14)} is given by
\begin{align}
   \Tilde{B}(b)&\equiv\sum_{n=0}^\infty\frac{\Tilde{k}_n}{n!}\,b^n
\notag\\
   &=(2e^{5/3}t\mu^2)^{\beta_0b/(4\pi)}\mathit{\Gamma}(2-\beta_0b/(4\pi)).
\label{eq:(2.15)}
\end{align}
The singularities of the Borel transform are located
at~$u\equiv\beta_0b/(4\pi)=2$, $3$, $4$, \dots, while the so-called ultraviolet (UV) renormalons
(singularities at negative $b$) do not exist.
This is because the UV
behavior is improved by the gradient flow.

\section{Comparison of the large-$\beta_0$ approximation and the explicit
perturbative computation for $\hat{E}(t^{-1})$}
\label{sec:C}
It is interesting to assess the quality of the large-$\beta_0$ approximation for $\hat{E}(t^{-1})$.
We compare the results in the large-$\beta_0$ approximation computed in Appendix~\ref{sec:A} with the explicit perturbative calculation
in~Refs.~\cite{Luscher:2010iy,Harlander:2016vzb}. Defining the perturbative series as
\begin{equation}
   \hat{E}(t^{-1})
   =\frac{3\dim(G)}{32\pi}\alpha
   \left(1+k_1\alpha+k_2\alpha^2+\mathcal{O}(\alpha^3)\right),
\label{eq:(B1)}
\end{equation}
one has~\cite{Luscher:2010iy}
\begin{equation}
   k_1=\frac{1}{4\pi}\beta_0L
   +\frac{1}{4\pi}
   \left[\left(\frac{11}{3}\gamma_E+\frac{52}{9}-3\ln3\right)C_A
   +\left(-\frac{4}{3}\gamma_E-\frac{8}{9}+\frac{8}{3}\ln2
   \right)T_R n_f\right],
\label{eq:(B3)}
\end{equation}
and~\cite{Harlander:2016vzb}
\begin{align}
   k_2&=\frac{1}{(4\pi)^2}\beta_0^2L^2+\frac{1}{(4\pi)^2}\beta_1L
\notag\\
   &\qquad{}
   +\frac{2}{(4\pi)^2}\beta_0
   \left[\left(\frac{11}{3}\gamma_E+\frac{52}{9}-3\ln3\right)C_A
   +\left(-\frac{4}{3}\gamma_E-\frac{8}{9}+\frac{8}{3}\ln2
   \right)T_R n_f\right]L
\notag\\
   &\qquad\qquad{}
   +8
   \bigl\{
   -0.013\,642\,3(7)C_A^2
\notag\\
   &\qquad\qquad\qquad\qquad{}
   +\left[
   0.006\,440\,134(5)C_F-0.008\,688\,4(2)C_A
   \right]T_R n_f
\notag\\
   &\qquad\qquad\qquad\qquad\qquad{}
   +0.000\,936\,117T_R^2n_f^2
   \bigr\} \, ,
\label{eq:(B4)}
\end{align}
where we set
\begin{equation}
   L\equiv\ln(8\mu^2 t) \, .
\label{eq:(B2)}
\end{equation}
Here $\beta_0$ is given by~Eq.~\eqref{eq:(2.11)} and $\beta_1$ is the two-loop
coefficient of the beta function,
\begin{equation}
   \beta_1\equiv
   \frac{34}{3}C_A^2-\left(4C_F+\frac{20}{3}C_A\right)T_R n_f.
\label{eq:(B5)}
\end{equation}

\begin{figure}[t!]
\begin{minipage}{0.5\hsize}
\begin{center}
\includegraphics[width=70mm]{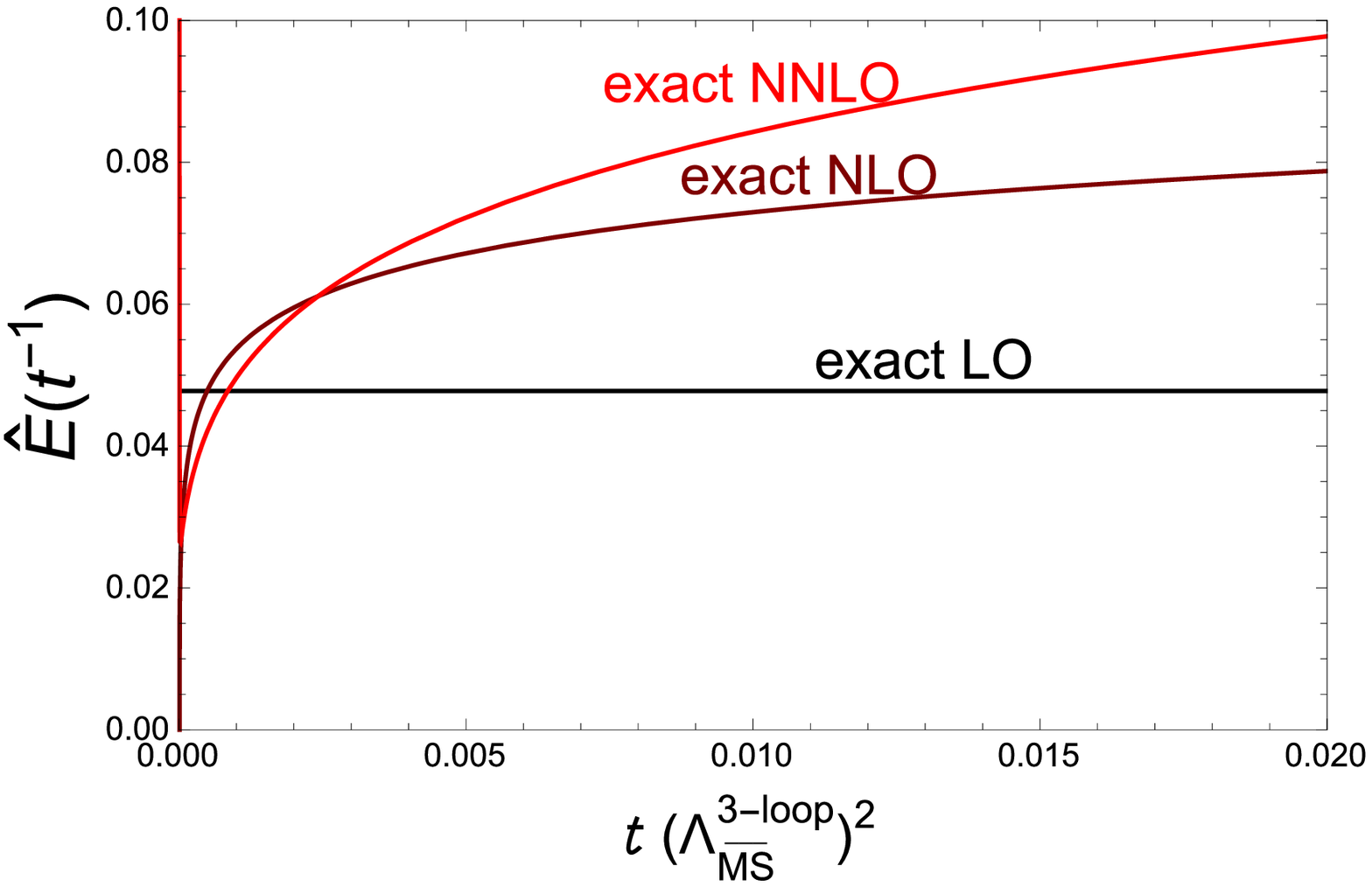}
\end{center}
\end{minipage}
\begin{minipage}{0.5\hsize}
\begin{center}
\includegraphics[width=70mm]{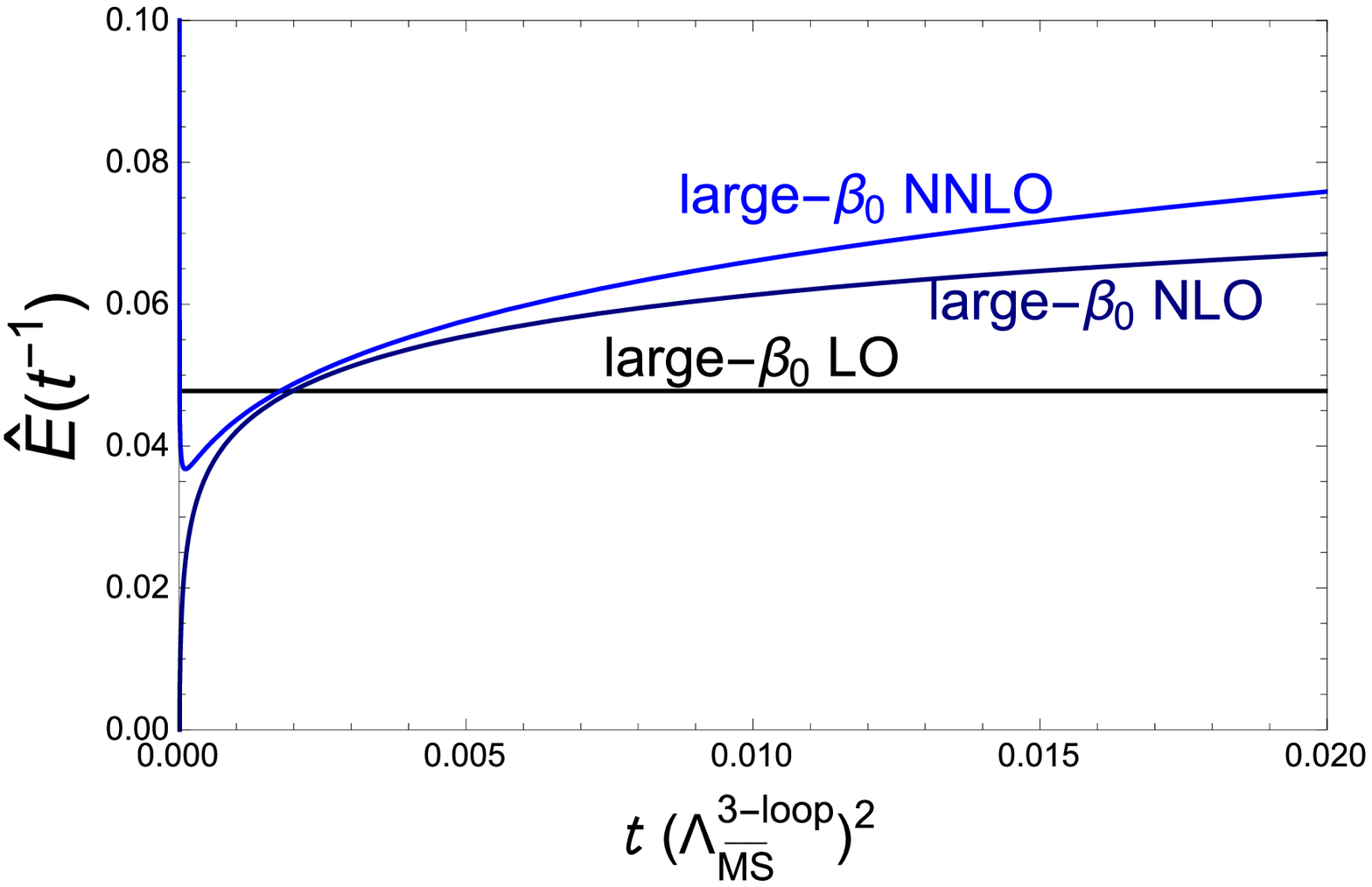}
\end{center}
\end{minipage}
\caption{Perturbative series for $\hat{E}(t^{-1})$: exact result (left) and large-$\beta_0$ approximation (right).
The N$^k$LO line represents the sum of the series up to $\mathcal{O}(\alpha^{k+1})$.
The input $\alpha(\mu)=0.2$ is used, which corresponds to $( \Lambda^{\text{ 3-loop}}_{\rm{ \overline{MS}}} )^2/\mu^2 \simeq 0.014$.
We set $G=SU(3)$ and $n_f=0$.}
\label{fig:exactvslargebeta}
\end{figure}

The perturbative coefficients in the large-$\beta_0$
approximation~\eqref{eq:(2.14)}, on the other hand, are obtained as
\begin{equation}
   \Tilde{k}_1=\frac{1}{4\pi}\beta_0L
   +\frac{1}{4\pi}
   \left[\left(\frac{11}{3}\gamma_E+\frac{22}{9}-\frac{22}{3}\ln2\right)C_A
   +\left(-\frac{4}{3}\gamma_E-\frac{8}{9}+\frac{8}{3}\ln2\right)T_Rn_f
   \right],
\label{eq:(B6)}
\end{equation}
and
\begin{align}
   \Tilde{k}_2&=\frac{1}{(4\pi)^2}\beta_0^2L^2
\notag\\
   &\qquad{}
   +\frac{2}{(4\pi)^2}\beta_0
   \left[\left(\frac{11}{3}\gamma_E+\frac{22}{9}-\frac{22}{3}\ln2\right)C_A
   +\left(-\frac{4}{3}\gamma_E-\frac{8}{9}+\frac{8}{3}\ln2\right)T_R n_f\right]L
\notag\\
   &\qquad\qquad{}
   +\frac{1}{(4\pi)^2}
   \left[\frac{11}{3}C_A-\frac{4}{3}T_Rn_f\right]^2
\notag\\
   &\qquad\qquad\qquad\qquad\qquad{}
   \times\left(
   \gamma_E^2+\frac{4}{3}\gamma_E-4\gamma_E\ln2
   -\frac{5}{9}+\frac{\pi^2}{6}+4\ln^22-\frac{8}{3}\ln2
   \right)
\notag\\
   &=\frac{1}{(4\pi)^2}\beta_0^2L^2
\notag\\
   &\qquad{}
   +\frac{2}{(4\pi)^2}\beta_0
   \left[\left(\frac{11}{3}\gamma_E+\frac{22}{9}-\frac{22}{3}\ln2\right)C_A
   +\left(-\frac{4}{3}\gamma_E-\frac{8}{9}+\frac{8}{3}\ln2\right)T_R n_f\right]L
\notag\\
   &\qquad\qquad{}
   +8
   \left[0.007\,079\,38C_A^2
   -0.005\,148\,64C_A T_R n_f
   +0.000\,936\,117T_R^2n_f^2\right].
\label{eq:(B7)}
\end{align}
From the above expressions, 
we can confirm that the
leading logarithmic terms [i.e., the $\mathcal{O}(L)$~term in~$k_1$ and the
$\mathcal{O}(L^2)$~term in~$k_2$] are correctly reproduced in the
large-$\beta_0$ approximation (which is a general feature of the large-$\beta_0$ approximation).
Also, we see that the leading large-$n_f$ terms,
the $\mathcal{O}(n_f)$~term in~$k_1$ and the $\mathcal{O}(n_f^2)$~term
in~$k_2$, are correctly reproduced; this is also expected because the
large-$\beta_0$ approximation becomes exact in the large-$n_f$ limit.

We now compare the behavior of the perturbative series obtained in the large-$\beta_0$ approximation
with that in the exact calculations. In Fig.~\ref{fig:exactvslargebeta}, we show the result for $G=SU(3)$ and $n_f=0$.
Since the perturbative coefficients in the large-$\beta_0$ approximation used here contain the parts which are not generally reproduced correctly,
this is a non-trivial check of the quality of the large-$\beta_0$ approximation.
One sees that they have qualitatively similar behavior.

\section{Attempt at a numerical estimate of the gluon condensate}
\label{sec:D}
In this section, we attempt a numerical estimate of the renormalon-free gluon condensate~Eq.~\eqref{RFglu}.
For this, we use lattice data of $\hat{E}(t^{-1})$.
We compare it with the renormalon-free OPE formula given in Eq.~\eqref{EtOPE}
to extract the value of the gluon condensate, using $c_{\mathbbm{1}, \hat{E}}^{\rm RF}(t)$
given in Eq.~\eqref{ERF}.
We exhibit how well (or not) our framework works at a practical level, which is based on the large-$\beta_0$ approximation.

\begin{figure}[t]
\centering
\includegraphics[width=0.5\columnwidth]{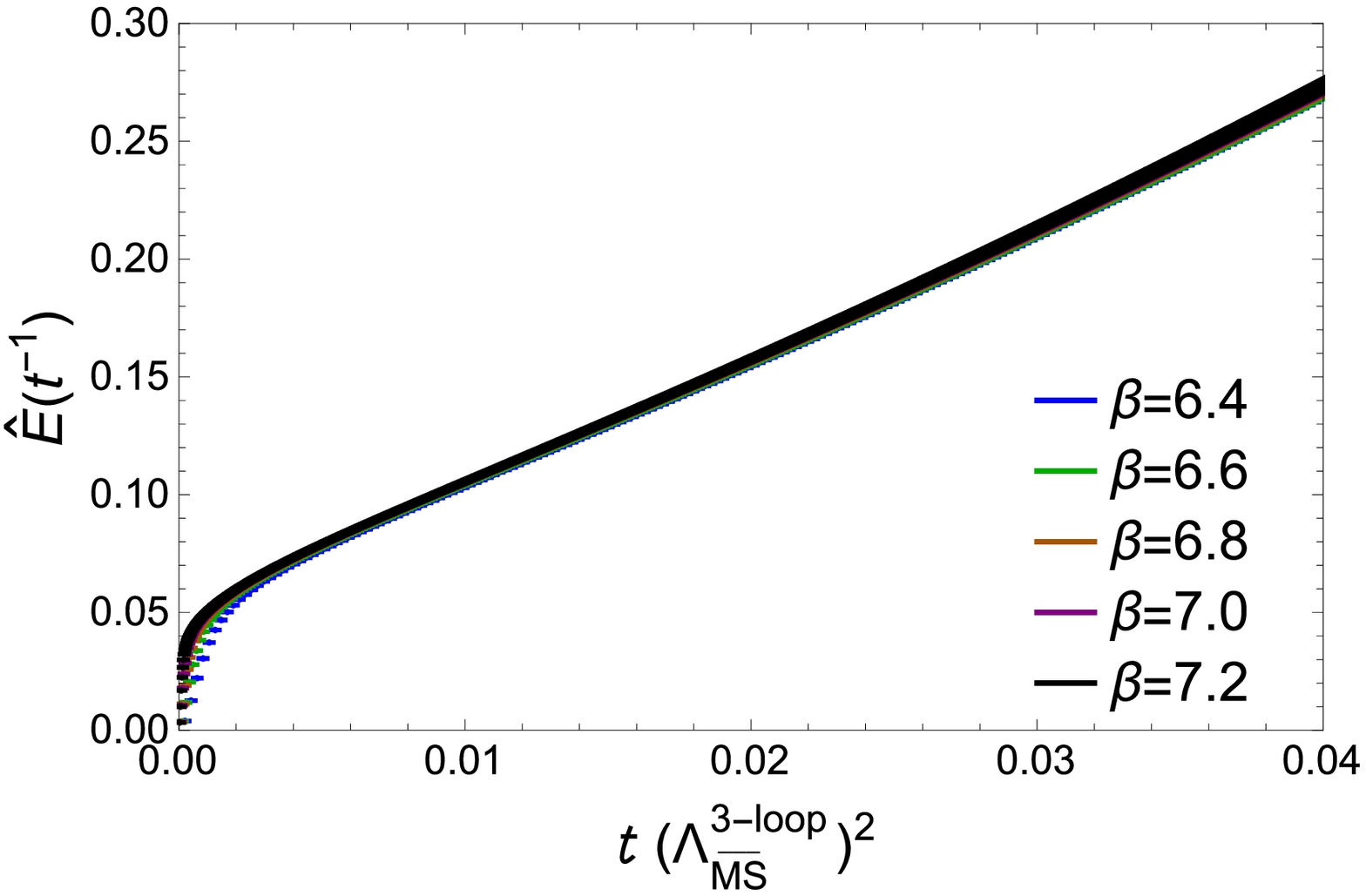}
\caption{Lattice results for~$\hat{E}(t^{-1})$. Different colored lines
correspond to different~$\beta$. The statistical error is represented by the width of the line.}
\label{fig:4}
\end{figure}

We use lattice data obtained by the FlowQCD
collaboration~\cite{Asakawa:2015vta,Kitazawa:2016dsl}.\footnote{We are grateful
to Masakiyo Kitazawa for providing us with the numerical data.} In~Fig.~\ref{fig:4},
we show the lattice results for~$\hat{E}(t^{-1})$ for the bare gauge
couplings, $\beta=6.4$, $6.6$, $6.8$, $7.0$, and~$7.2$.
To show the lattice data in~$\Lambda_{\overline{\text{MS}}}^{\text{3-loop}}$ units,
we used the relation between $\beta$ and the lattice spacing~$a$ obtained
in~Ref.~\cite{Kitazawa:2016dsl}.\footnote{We neglect the estimated errors
in~Ref.~\cite{Kitazawa:2016dsl} in our analysis.} We see that the lattice data
among different $\beta$'s overlap each other in the region
$t(\Lambda_{\overline{\text{MS}}}^{\text{3-loop}})^2\gtrsim0.01$. Therefore, we use
the lattice data at~$t(\Lambda_{\overline{\text{MS}}}^{\text{3-loop}})^2\geq0.01$ of the finest lattice spacing ($\beta=7.2$
and~$a^2(\Lambda_{\overline{\text{MS}}}^{\text{3-loop}})^2=5.3\times10^{-4}$, shown
by the black line in~Fig.~\ref{fig:4}) regarding it as
the continuum limit.\footnote{The selected region
$t(\Lambda_{\overline{\text{MS}}}^{\text{3-loop}})^2\geq0.01$ satisfies
$\sqrt{t}\gtrsim4a$. We adopt such a large scale hierarchy to suppress the
finite $a$ effect, taking into account that we do not take the continuum limit.}

We compare the lattice result in Fig.~\ref{fig:4} with the renormalon-free part $c_{\mathbbm{1}, \hat{E}}^{\rm RF}(t^{-1})$. 
To compare them quantitatively,
we need the ratio $r\equiv\Lambda_{\overline{\text{MS}}}^{\text{1-loop}}/
\Lambda_{\overline{\text{MS}}}^{\text{3-loop}}$, because our theoretical calculation
on the basis of the large-$\beta_0$ approximation is given
in~$\Lambda_{\overline{\text{MS}}}^{\text{1-loop}}$ units whereas the lattice results
are shown in~$\Lambda_{\overline{\text{MS}}}^{\text{3-loop}}$ units. We determine
this ratio by requiring the running couplings at one-loop and three-loop to have the
same value at $\mu=a^{-1}$, i.e. we impose
$\alpha_{s,{\text{1-loop}}}(a^{-1}/\Lambda_{\overline{\text{MS}}}^{\text{1-loop}})=
\alpha_{s,{\text{3-loop}}}(a^{-1}/\Lambda_{\overline{\text{MS}}}^{\text{3-loop}})=
0.1214$. (Note that $\alpha_{s, \text{3-loop}}$ at this scale is determined from $a^2(\Lambda_{\overline{\text{MS}}}^{\text{3-loop}})^2=5.3\times10^{-4}$
since the running coupling at $k$-loop is a function of~$\mu/\Lambda_{\overline{\text{MS}}}^{\text{$k$-loop}}$.) This condition ensures
that the calculation at leading-log (LL) matches well with the one at
next-to-next-to-LL (NNLL) around the region
$t(\Lambda_{\overline{\text{MS}}}^{\text{3-loop}})^2\sim
a^{-2}(\Lambda_{\overline{\text{MS}}}^{\text{3-loop}})^2=
5.3\times10^{-4}$.\footnote{The LL prediction is the one-loop renormalization
group (RG) improvement of the leading-order (LO) prediction. Similarly, the
NNLL prediction is the three-loop RG improvement of the NNLO prediction. Due to the
matching of the coupling at the lattice cutoff, the difference between these
predictions at~$t(\Lambda_{\overline{\text{MS}}}^{\text{3-loop}})^2\sim
a^{-2}(\Lambda_{\overline{\text{MS}}}^{\text{3-loop}})^2=
5.3\times10^{-4}$ is of order $\alpha^2\sim0.12^2$.}
This is legitimate because both predictions should be accurate in such a short-distance region. ]
The above condition yields~$r=0.395$.

\begin{figure}[tp]
\centering
\begin{subfigure}{0.45\columnwidth}
\centering
\includegraphics[width=\columnwidth]{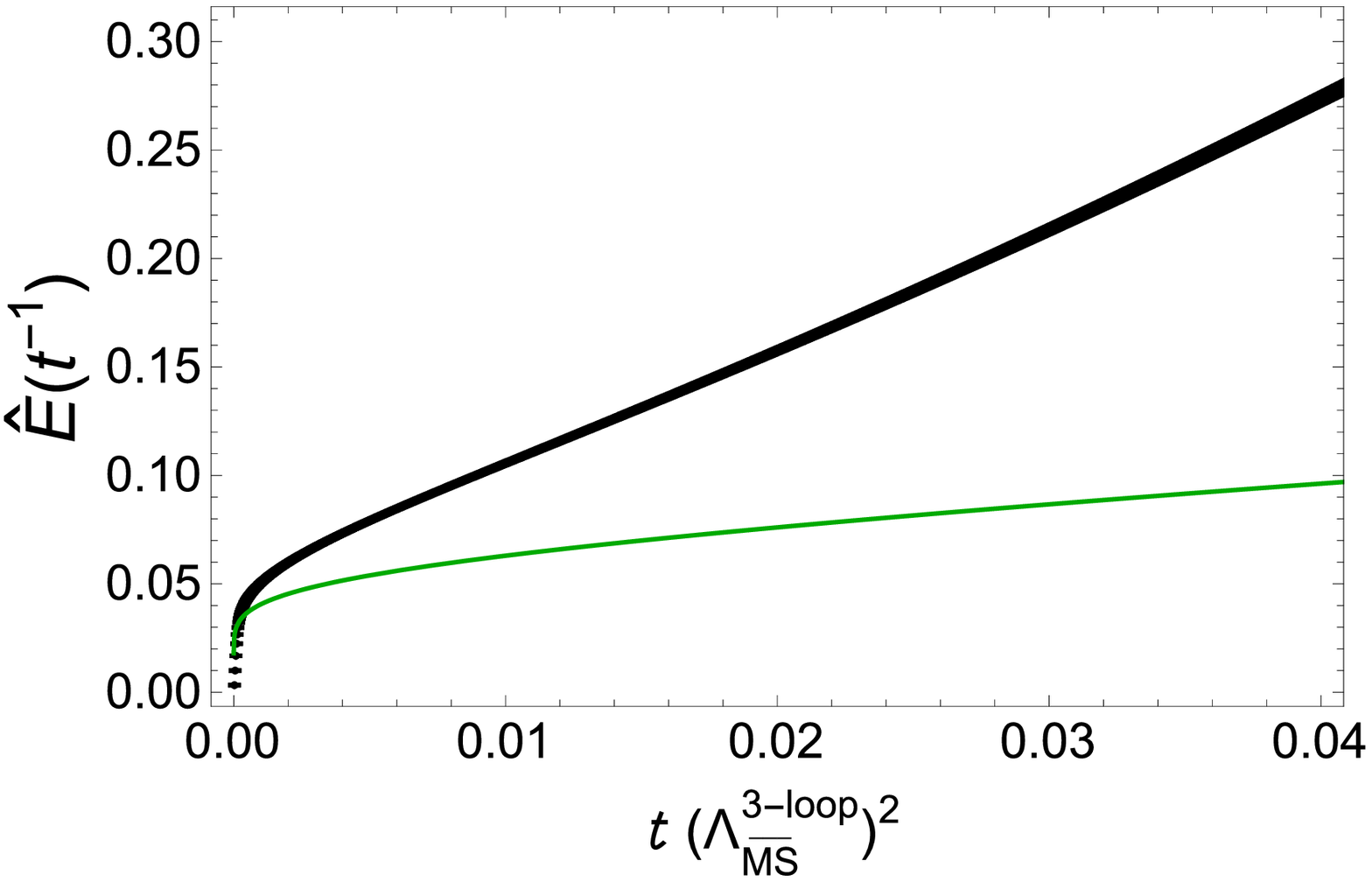}
\end{subfigure}
\hspace*{2em}
\begin{subfigure}{0.45\columnwidth}
\centering
\includegraphics[width=\columnwidth]{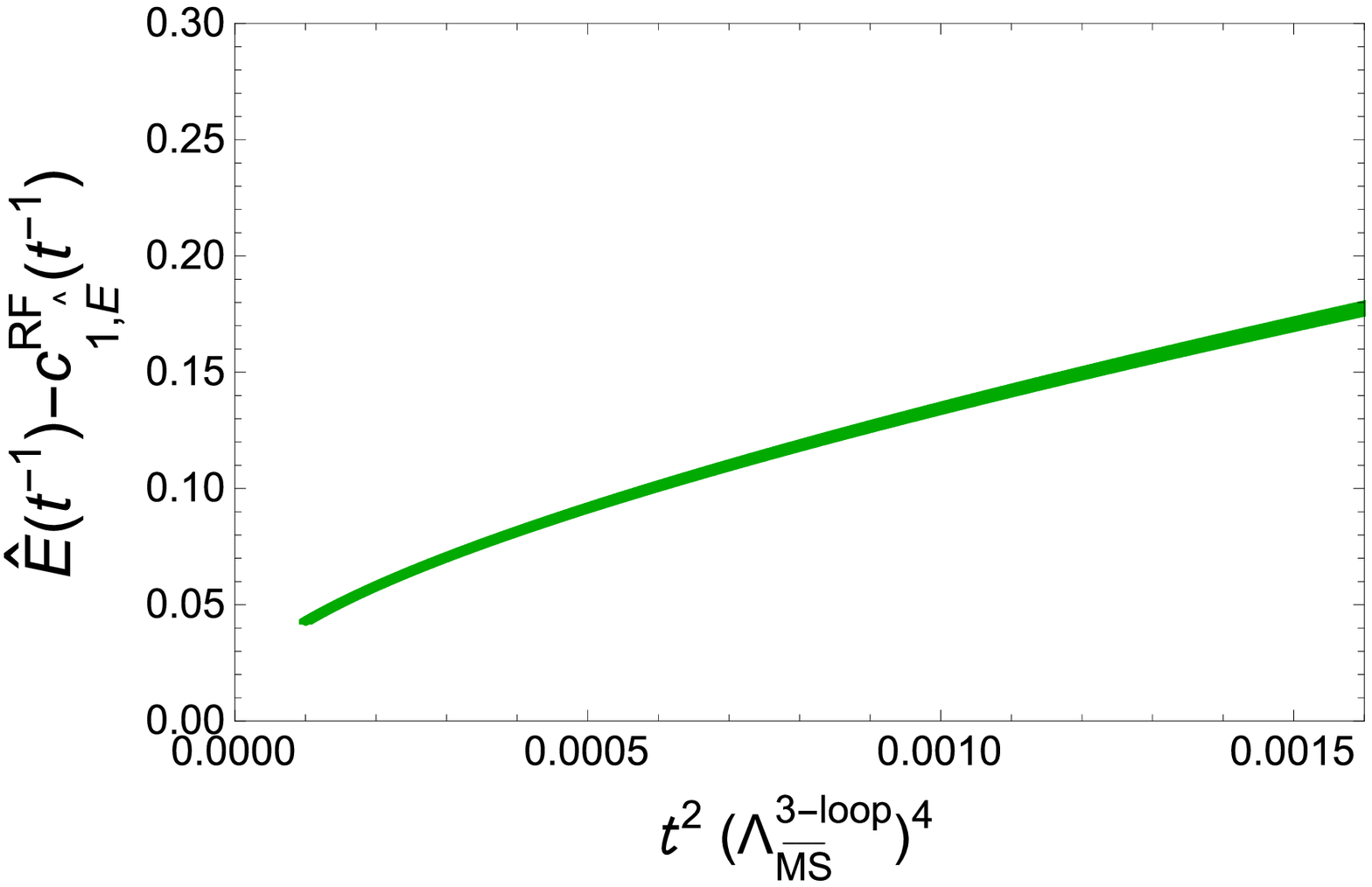}
\end{subfigure}
\centering
\includegraphics[width=0.45\columnwidth]{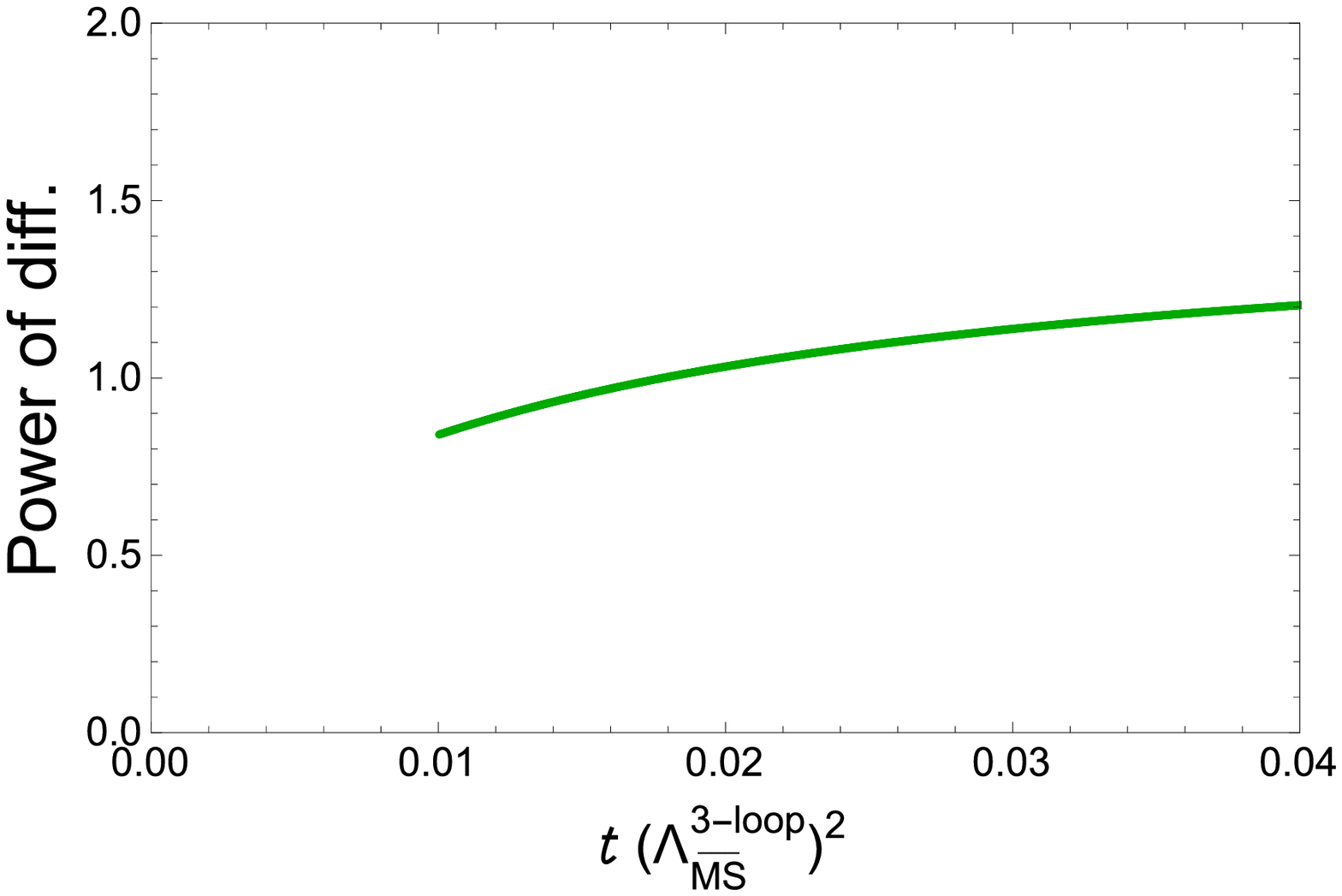}
\caption{Comparison of~$c_{\mathbbm{1}, \hat{E}}(t^{-1})$~\eqref{ERF}
with the lattice result. In the upper left panel,
$c_{\mathbbm{1}, \hat{E}}(t^{-1})$ (green line) and the lattice data (black
line) are shown together as functions
of~$t(\Lambda_{\overline{\text{MS}}}^{\text{3-loop}})^2$. The difference between
them is shown in the upper right panel, where the horizontal axis is taken as
$t^2(\Lambda_{\overline{\text{MS}}}^{\text{3-loop}})^4$ in order to examine whether the
linear behavior expected from Eq.~\eqref{EtOPE} is observed. In the lower
panel, we show an effective power of the difference
in~$t(\Lambda_{\overline{\text{MS}}}^{\text{3-loop}})^2$. Statistical error is not
estimated in this last figure. In the upper right and lower panels, we show only the
data points in the
region~$t(\Lambda_{\overline{\text{MS}}}^{\text{3-loop}})^2\gtrsim0.01$.}
\label{fig:5}
\end{figure}

Using this~$r$, in~Fig.~\ref{fig:5} we compare the
lattice result with the renormalon-free part, $c_{\mathbbm{1},\hat{E}}^{\rm RF}$ in~Eq.~\eqref{ERF}.
The difference between them, shown in the upper right panel, is expected to
have a linear behavior in~$t^2(\Lambda_{\overline{\text{MS}}}^{\text{3-loop}})^4$
according to the OPE or small flow time expansion~Eq.~\eqref{EtOPE}. To investigate quantitatively if this is the
case or not, in the lower panel, we plot an effective power of the difference
defined by~$d\ln f(x)/d(\ln x)$, where $f(x)$ is the difference and
$x\equiv t(\Lambda_{\overline{\text{MS}}}^{\text{3-loop}})^2$. From the lower panel,
it seems that a component with the power smaller than~2 remains in the
difference, i.e., the difference does not show $t^2$ behavior. 
Thus, we cannot extract the gluon condensate,
which is the coefficient of the $t^2$ term of the OPE \eqref{EtOPE}.

This failure is attributed to the fact that we use the large-$\beta_0$ approximation to evaluate the Wilson coefficient $c_{\mathbbm{1},\hat{E}}(t^{-1})$.
In this approximation, the perturbative error does not reach its minimal error (renormalon uncertainty) of~$\sim t^2$,
which is expected to be observed in sufficiently large-order perturbative calculations. 
This is not surprising because the large-$\beta_0$ approximation
takes into account the partial set of the Feynman diagrams and is accurate only at the LL level.
In case we do not know a sufficiently large-order result, the difference between nonperturbative (lattice) and perturbative results behaves as $\sim \alpha^{n}(1/\sqrt{t})$
rather than $t^2$.

\begin{figure}[tbp]
\centering
\begin{subfigure}{0.45\columnwidth}
\centering
\includegraphics[width=\columnwidth]{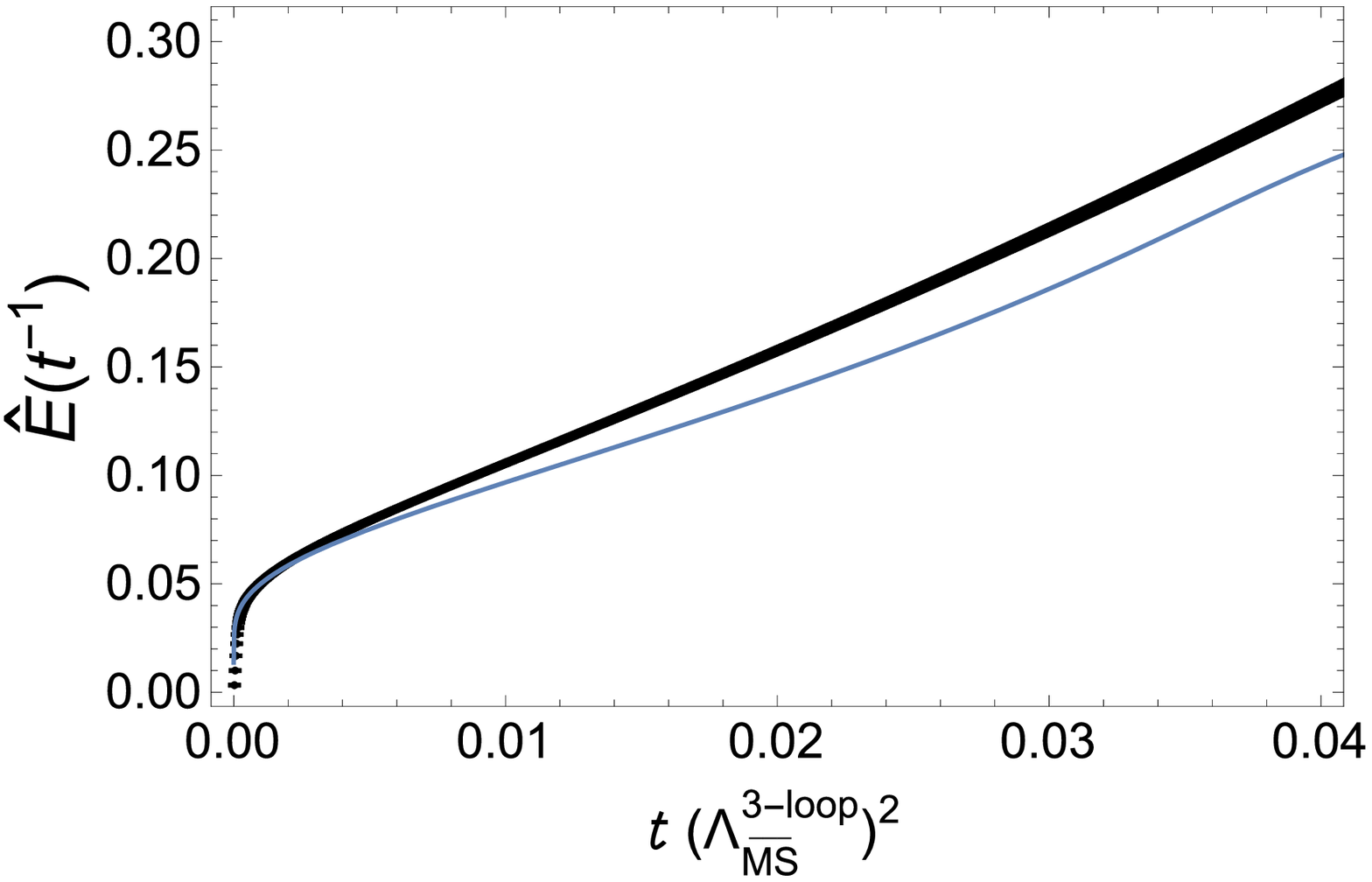}
\end{subfigure}
\hspace*{2em}
\begin{subfigure}{0.45\columnwidth}
\centering
\includegraphics[width=\columnwidth]{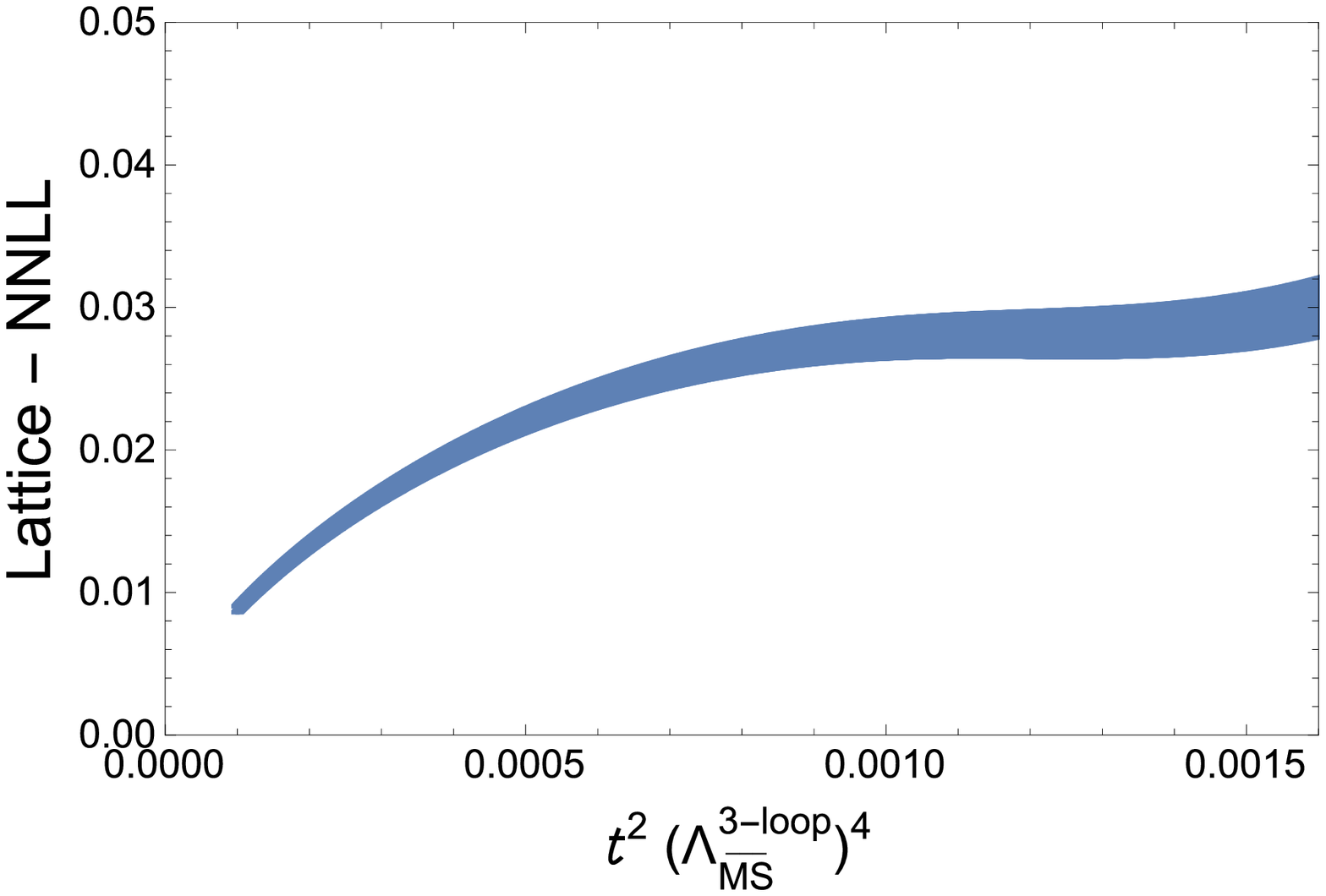}
\end{subfigure}
\centering
\includegraphics[width=0.55\columnwidth]{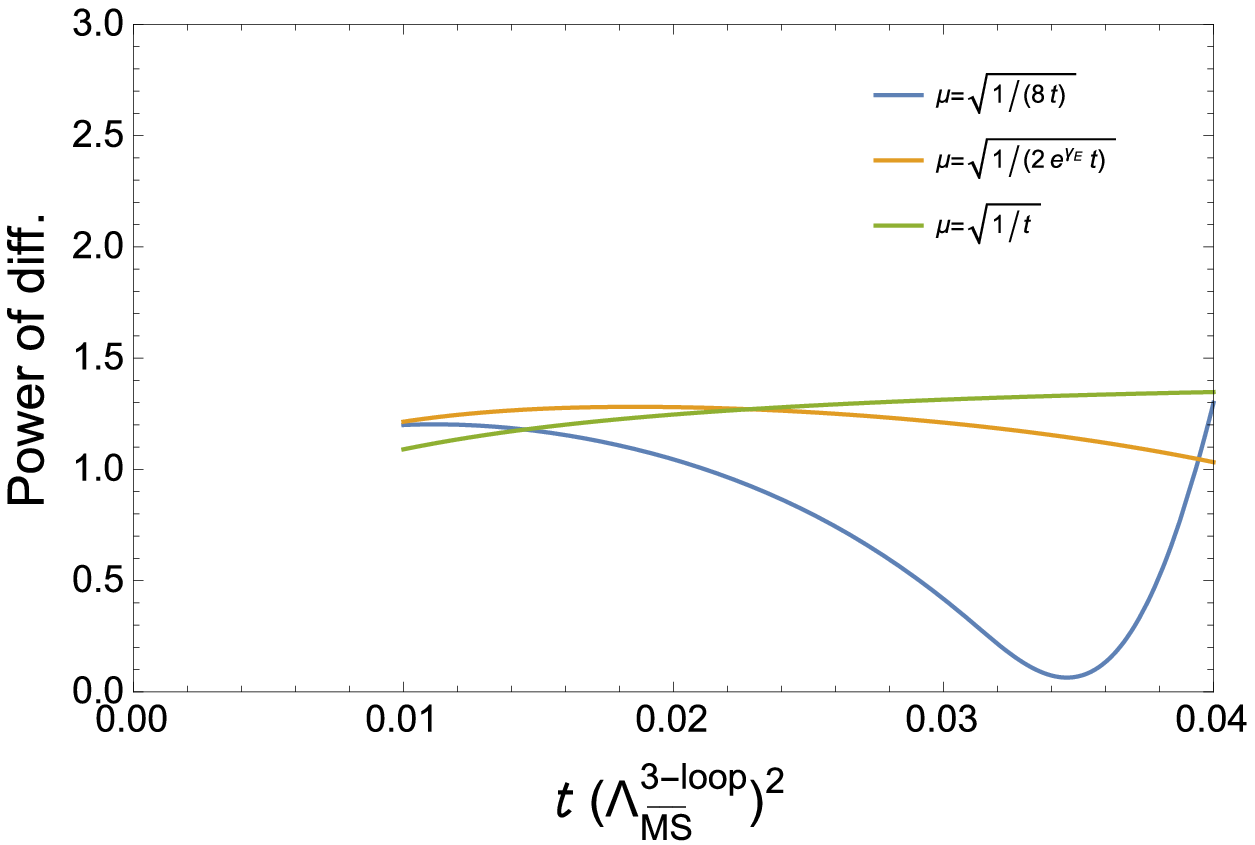}
\caption{Comparison of~NNLL result with the lattice result. See the caption of Fig.~\ref{fig:5}.}
\label{fig:6}
\end{figure}

Although the large-$\beta_0$ approximation is not sufficient to detect $t^2$ behavior,
we now investigate whether such behavior is observed when we use the exact perturbative calculation,
which is currently known up to NNLO, namely $\mathcal{O}(\alpha^3)$ \cite{Harlander:2016vzb}. 
In Fig.~\ref{fig:6} we compare the NNLL result with the lattice result.
The renormalization scale is taken as $\mu=1/\sqrt{8 t}$.
We again examine the effective power in $x=t (\Lambda_{\overline{\text{MS}}}^{\text{3-loop}})^2$ of their difference,
which turns out to still be smaller than 2.
We also show the results with different choices of the renormalization scale
but they exhibit similar results.

From the above analyses, we conclude that in order to determine the renormalon-free gluon condensate reliably,
we need a formulation beyond the large-$\beta_0$ approximation 
and also need further higher-order results than are currently available.

\end{document}